\documentclass[11pt,a4paper]{article}
\usepackage[margin=2.5cm]{geometry}
\usepackage{amsmath,amssymb,amsthm}
\usepackage{stmaryrd}
\newcommand{\NT}[1]{\llbracket #1\rrbracket}
\usepackage{enumitem}
\usepackage[hidelinks]{hyperref}
\newcommand{\fwdlink}[2]{%
  \hyperlink{#1}{#2\textsuperscript{\(\nearrow\)}}%
}
\newtheorem{theorem}{Theorem}[section]
\newtheorem{lemma}[theorem]{Lemma}
\newtheorem{proposition}[theorem]{Proposition}
\newtheorem{corollary}[theorem]{Corollary}
\newtheorem{definition}[theorem]{Definition}
\theoremstyle{remark}
\newtheorem{remark}[theorem]{Remark}
\theoremstyle{plain}
\newcommand{\D}{\mathcal{D}}
\newcommand{\WitnessSet}{\mathsf{Wit}}
\newcommand{\cf}{\mathrm{cf}}
\newcommand{\lin}{\mathrm{lin}}
\newcommand{\Ccf}[1]{\mathcal C_{#1}^{\cf}}
\newcommand{\Clin}[1]{\mathcal C_{#1}^{\lin}}
\newcommand{\pref}{\mathrm{pref}}
\newcommand{\suf}{\mathrm{suf}}
\title{Distributional Learning of Context-Free Languages under Fixed Finite-Semigroup Typing}
\author{Takayuki Kuriyama\\
Independent Researcher, Tokyo, Japan\\
\texttt{growup.kuriyama@gmail.com}}
\date{}
\begin{document}
\maketitle

\begin{abstract}
We study positive-data learning of context-free languages whose nonempty
factors are typed by a fixed morphism $s:\Sigma^+\to\mathsf S$ into a finite
semigroup.  A polynomial-time set-driven operator reconstructs each
substitutable target from finite positive witnesses; a separate conservative
learner identifies nonempty targets in Gold's sense.  The encoded
characteristic-data size is polynomial in grammar size and a typed
shortest-yield parameter.  Ordinary grammar thickness suffices when the
typing is determined by a fixed prefix--suffix window and a finite-group
value; for linear targets, linear grammar size alone suffices.  Conversely,
a fixed two-element semilattice typing yields finite targets with
grammars of size and ordinary thickness both $O(n)$ that require exponentially long
characteristic samples.  The lower bound applies to any set-driven
operator admitting characteristic samples for the fixed-typing class.
Retyping closure, examples beyond fixed windows, and a Dyck-language
obstruction delineate the framework's scope.
\end{abstract}

\par\medskip

\section{Introduction}

\subsection{Background and motivation}

Gold~\cite{gold1967} established limitations of positive-data
identification. Distributional learning yields positive results
for restricted context-free classes, notably Clark--Eyraud
substitutability~\cite{clark-eyraud2007} and Yoshinaka's
$(k,\ell)$-substitutability~\cite{yoshinaka2008}.
For background see de~la~Higuera~\cite{higuera2010};
Kanazawa's work on learning categorial grammars offers a
complementary approach~\cite{kanazawa1996,kanazawa1998}.

\paragraph{Relation to earlier work.}
The author's 2014 preprint~\cite{kuriyama2014} introduced
substitutability relative to recognizable equivalence relations,
preceding the publication of Yoshinaka's general many-sorted
signature framework~\cite{yoshinaka2015}.
For a fixed semigroup morphism $s:\Sigma^+\to\mathsf S$,
its nonempty fibers can be represented as disjoint sorts,
equipped with typed concatenation and promotion to the
whole-string sort. Under this specialization, Yoshinaka's
general framework~\cite[Theorem~1 and Lemma~3]{yoshinaka2015}
already supplies positive-data identification, polynomial-time
hypothesis updates, and characteristic witnesses of polynomial
cardinality in a typed target grammar. The particular choice
of sorts is not attributed to that work.

The procedures below give an explicit fixed-typing construction,
including a direct completeness proof for nullable and unit
productions. Their existence, update-time bound, and finite-witness
cardinality are not claimed as independent learning-theoretic
novelties. Our principal new questions concern the \emph{encoded
length} of characteristic data, its polynomial upper and
exponential lower bounds, and the algebraic and expressive
consequences of finite typing. The locally trivial fixed-window
criterion itself is classical.

We study a finite-semigroup-congruence version of this program.  Let
$s:\Sigma^+\to\mathsf S$ be a homomorphism into a finite semigroup and
$\sim_s:=\ker s\subseteq\Sigma^+\times\Sigma^+$.  For $x\in\Sigma^*$, write
$\D_L(x):=\{(u,v):uxv\in L\}$.  A language $L$ is
\emph{$\sim_s$-substitutable} if, for nonempty $x,y$,
$s(x)=s(y)$ and $\D_L(x)\cap\D_L(y)\ne\emptyset$ imply
$\D_L(x)=\D_L(y)$. Write $\Ccf{s}$ for the context-free
$\sim_s$-substitutable languages, and call this the \emph{fixed-$s$ slice}.
Thus full distributional equality is required only after a fixed finite
algebraic comparison.

For a minimal example, take $\Sigma=\{a\}$ and $L=\{a,aa\}$.
Although $(\lambda,\lambda)\in\D_L(a)\cap\D_L(aa)$,
the context $(a,\lambda)$ belongs to $\D_L(a)$ but not to $\D_L(aa)$.
Thus $L$ is not Clark--Eyraud substitutable~\cite[Section~6.1]{clark-eyraud2007}.
Under the parity morphism $s(a^n)=n\bmod 2$ into
$\mathbb Z/2\mathbb Z$, however, $a$ and $aa$ have distinct types
and $L\in\Ccf{s}$.

Finite-index semigroup congruences on $\Sigma^+$,
equivalently kernels of finite-semigroup homomorphisms,
provide finite typings.  For each fixed $s$,
write $\mathsf{RS}_s$ for the class of all $\sim_s$-substitutable
languages, without a context-freeness restriction.  For expressive
comparison, let
$\mathsf{RS}:=\bigcup_s\mathsf{RS}_s$,
where $s$ ranges over finite-semigroup homomorphisms on $\Sigma^+$.
Every regular language belongs to this union.  Our learning results,
however, concern only $\Ccf{s}$ for one fixed $s$; no learner is
claimed for $\mathsf{RS}$.

Unlike \fwdlink{def:kl-subst}{one fixed prefix--suffix window}, $s$ can combine independent
finite-state factor features by a product.  If the finite typing is allowed
to be refined accordingly, the framework is stable under regular filtering
and arbitrary inverse homomorphisms, including erasing ones
(Proposition~\ref{prop:finite-info-closure}).  This is a
change-of-typing statement, not closure of one fixed-$s$ slice.  An algebraic criterion identifies the typings determined by fixed
prefix--suffix windows on nonempty factors
(Proposition~\ref{prop:li-window}); combining windows with finite groups
yields a broader characteristic-data bound (Corollary~\ref{cor:li-thickness}).
Examples below exhibit nonregular linear and nonlinear targets
outside every fixed-window class, while counter and Dyck obstructions
delimit the reach of fixed finite-semigroup typing.

The choice of $s$ supplies a typing bias on terminal
factors~\cite{coste-typing2004}, not their occurrences.
Its connection with regular control
sets~\cite{takada1988,takada1995} is confined to
the linear setting of Section~\ref{sec:linear}.

\subsection{Contributions and main results}

The finite-semigroup typing $s$ is fixed as prior information
for the whole target class $\Ccf{s}$; it is not inferred
from positive data. We give a set-driven reconstruction operator
$\mathsf{Rec}_s$ and a separate conservative Gold learner
$\mathsf{Learn}_s$. Their qualitative learning and
polynomial-time guarantees specialize Yoshinaka's general
framework~\cite{yoshinaka2015}, using the substring-indexed
grammar construction of~\cite{yoshinaka2008}, and are not
claimed as independent novelties. Our characteristic-data
bounds concern $\mathsf{Rec}_s$; no incremental
polynomial-data guarantee is claimed for $\mathsf{Learn}_s$.
Our contributions are the following.

\begin{enumerate}[label=\textup{(\arabic*)},leftmargin=2.4em]

\item \emph{The quantitative cost of finite typing.}
For arbitrary fixed $s$, typed shortest-yield lengths
in a normalized grammar control the encoded size of
characteristic data
(Corollary~\ref{cor:typed-thickness-data}).
When $s$ is determined by a fixed prefix--suffix window
and a finite-group value, the size and ordinary shortest-yield
thickness of the original reduced CFG suffice
(Corollary~\ref{cor:li-thickness}).
Equivalently, the positive image $s(\Sigma^+)$ may be any
homomorphic image of a subsemigroup of a product of a
finite locally trivial semigroup and a finite group
(the class $\mathbf{LI}\vee\mathbf{Grp}$).
For linear targets, the size of an arbitrary linear CFG
alone suffices (Theorem~\ref{thm:linear-poly}).
The fixed-window case already admits polynomial
encoded-size bounds due to
Yoshinaka~\cite[Theorem~1]{yoshinaka2008}.

In contrast, a fixed two-element semilattice typing yields
finite targets $L_n$ with reduced grammars of size and
ordinary thickness $O(n)$ for which every characteristic
sample must contain a word $w_n^*$ of length $3\cdot2^n-2$.
This holds for every set-driven operator admitting
characteristic samples for both $L_n$ and
$L_n^-:=L_n\setminus\{w_n^*\}$, which also belongs
to the same fixed-typing class
(Proposition~\ref{prop:typed-thickness-gap}).
The targets $L_n$ admit polynomial characteristic data
under trivial typing.

\item \emph{Expressive scope and limitations.}
We construct nonregular linear
(Proposition~\ref{prop:linear-separator-example})
and nonlinear
(Proposition~\ref{prop:nonlinear-rs-example})
targets outside every fixed-window class.
We show that the uncapped counter language
$\mathrm{CTR}$ and the one-bracket Dyck language
$D_1$ lie outside $\mathsf{RS}$
(Section~\ref{sec:outside-rs}).
We also place the context-free fixed-typing classes
within Clark's congruential family
(Proposition~\ref{prop:clark-congruential-comparison}).

\end{enumerate}

Elementary closure properties under changes of typing
and the classical characterization of fixed-window
typings are recorded in Section~\ref{sec:framework}.

\section{Preliminaries}
\label{sec:prelim}

Let $\Sigma$ be a finite alphabet, let $\Sigma^*$ be the free monoid with
identity $\lambda$, and put $\Sigma^+:=\Sigma^*\setminus\{\lambda\}$.
A pair $(u,v)\in\Sigma^*\times\Sigma^*$ is called a
\emph{terminal context}.  For $L\subseteq\Sigma^*$ and $x\in\Sigma^*$, the \emph{distribution} of $x$ in $L$ is
$\D_L(x):=\{(u,v)\in\Sigma^*\times\Sigma^*:uxv\in L\}$.
The \emph{syntactic congruence} of $L$ is defined by $x\equiv_L y$ iff
$\D_L(x)=\D_L(y)$, for $x,y\in\Sigma^*$~\cite[Chapter~III, Proposition~10.2]{eilenberg1974}.
As usual, $\equiv_L$ is a monoid congruence: if
$x\equiv_L x'$ and $y\equiv_L y'$, then $xy\equiv_L x'y'$.

For a word $w\in\Sigma^*$ and a letter $a\in\Sigma$, let $|w|_a$ denote
the number of occurrences of $a$ in $w$.

A context-free grammar (CFG) is a tuple $G=(V,\Sigma,P,S)$ with finite
nonterminal set $V$, start symbol $S\in V$, and productions
$A\to\alpha$ with $A\in V$ and $\alpha\in(V\cup\Sigma)^*$.
A \emph{sentential form} is any word in $(V\cup\Sigma)^*$.  We write
$\alpha\Rightarrow_G\beta$ for one derivation step in $G$, and
$\Rightarrow_G^*$ for its reflexive-transitive closure.  Thus $L(G):=\{w\in\Sigma^*:S\Rightarrow_G^*w\}$.  For $A\in V$, put
$L_G(A):=\{w\in\Sigma^*:A\Rightarrow_G^*w\}$.
For a nonterminal $X\in V$, a terminal context $(u,v)$ is a
\emph{terminal reaching context of $X$ in $G$} if
$S\Rightarrow_G^*uXv$.
A nonterminal is \emph{reachable} if it occurs in a sentential form reachable
from $S$, \emph{productive} if it derives a terminal word, \emph{useless}
if it is not both, and \emph{nullable} if it derives $\lambda$; $G$ is
\emph{reduced} if it has no useless nonterminal.  A \emph{unit production}
(\emph{unit rule}) is $A\to B$ with $A,B\in V$, and the \emph{unit closure}
is the reflexive-transitive closure of the induced relation on $V$.  For
$L(G)\ne\emptyset$, the \emph{trim} (\emph{reduced part}) of $G$ deletes
the nonproductive nonterminals and then the unreachable ones, each with its
incident productions.  A CFG is \emph{linear} if every right-hand side contains at
most one nonterminal.  By \emph{binarization} we mean replacement of longer
right-hand sides by equivalent chains of binary productions using fresh
nonterminals.
We write $\mathrm{CFL}$ for the class of context-free languages.  For any
finite representation $\mathcal R$, $|\mathcal R|$ denotes the length of a fixed reasonable
encoding; for a grammar this is polynomially equivalent to the usual symbol
count of its productions.  Following the thickness parameter used by
Wakatsuki--Tomita
\cite{wakatsuki-tomita1993} and adopted by Yoshinaka~\cite{yoshinaka2008}
(see also Eyraud--Heinz--Yoshinaka
\cite[Definition~2.11]{eyraud-heinz-yoshinaka2016}), for a reduced
CFG $G=(V,\Sigma,P,S)$ we use the shortest-yield \emph{thickness}
\[
 \tau_G:=\max_{A\in V}\min\{\,|w|:A\Rightarrow_G^*w,\ w\in\Sigma^*\,\}.
\]

\subsection{Learning from positive data}\label{sec:learning}

We use Gold's explanatory model~\cite{gold1967}.  Let $H_0$ denote a fixed
CFG with one start symbol and no productions, so $L(H_0)=\emptyset$.  We use
texts only for nonempty targets and represent the empty language separately by
$H_0$; equivalently one may adjoin the usual pause symbol.  For a nonempty language $L$, a \emph{text} is an
infinite sequence $\mathcal T=(w_1,w_2,\ldots)$ over $L$ in which every word of $L$
occurs.  Write $\mathcal T[n]=(w_1,\ldots,w_n)$ and
$K_n:=\{w_1,\ldots,w_n\}$.  A learner $\mathsf{Learn}$
\emph{identifies a nonempty $L$ in the limit} if for every text $\mathcal T$ for $L$
there are $n_0$ and a grammar $H$ such that $\mathsf{Learn}(\mathcal T[n])=H$ for every
$n\ge n_0$ and $L(H)=L$.
Thus the hypothesis representation itself eventually stabilizes.  A
sequential learner is \emph{conservative} if it changes its current hypothesis
only when the newly presented positive datum is not generated by that
hypothesis.

We distinguish this sequential learner from a \emph{set-driven reconstruction
operator} $\mathsf{Rec}$: this is a map from finite sets of words to grammars;
in particular, its output is independent of presentation order and multiplicity.
We call $\mathsf{Rec}$
\emph{sample-consistent} if $K\subseteq L(\mathsf{Rec}(K))$ for every finite
sample $K$.  A finite set $\mathcal F\subseteq L$ is a \emph{characteristic sample of $L$ for
$\mathsf{Rec}$} if
$\mathcal F\subseteq K\subseteq L$ implies $L(\mathsf{Rec}(K))=L$ for every finite
$K$; this is the positive-data, set-driven specialization of the
characteristic samples of de~la~Higuera~\cite{higuera1997}
(cf.\ \cite[Section~7.3.3, Definition~7.3.3]{higuera2010}, formulated for
sequential presentations).  We measure sample size by
$\|K\|:=\sum_{w\in K}(|w|+1)$, so that $\lambda$ counts and
$|K|\le\|K\|$, and call $\|\mathcal F\|$ the \emph{characteristic data} of $\mathcal F$.
Polynomial update time and polynomial characteristic data are distinct
criteria~\cite[Sections~7.3.2--7.3.3]{higuera2010}.  In the terms of
Eyraud--Heinz--Yoshinaka~\cite[Definitions~2.5 and~2.8]{eyraud-heinz-yoshinaka2016},
our polynomial characteristic-data guarantees are \emph{set-based}
(correct reconstruction by $\mathsf{Rec}_s$, defined in Section~4.1,
from every finite positive
sample $K$ satisfying $\mathcal F\subseteq K\subseteq L$, where
$\mathcal F$ is a characteristic sample for $L$); the \emph{incremental} criterion,
a correct and stable hypothesis of $\mathsf{Learn}_s$ (defined in Section~4.2)
as soon as such a sample
has appeared, is not claimed.  Gold stabilization of $\mathsf{Learn}_s$ is
proved in Corollary~\ref{cor:ilt}.

\subsection{Fixed finite-semigroup typing}\label{sec:fixed-typing}
Fix a finite semigroup $\mathsf S$ and a \emph{surjective}
semigroup homomorphism $s:\Sigma^+\twoheadrightarrow\mathsf S$,
represented by the multiplication table of $\mathsf S$
and the letter images $s(a)$ for $a\in\Sigma$.
Thus $s(w)$ is computed by a left-to-right product in
$O(|w|)$ time for nonempty $w$.
Surjectivity is without loss of generality: an arbitrary
finite semigroup typing may be replaced by its positive image.
Whenever a later construction produces a non-surjective
semigroup map (notably a product or inverse-image typing),
its type codomain is implicitly restricted to the positive
image before taking the fresh-identity completion.
We work over a nonempty alphabet unless stated otherwise;
for $\Sigma=\emptyset$, the only possible target languages
are $\emptyset$ and $\{\lambda\}$, which the unchanged learner
handles directly.

For a finite semigroup $\mathsf T$, an element
$e\in\mathsf T$ is \emph{idempotent} if $e^2=e$;
write $E(\mathsf T)$ for its idempotents.
Call $\mathsf T$ \emph{locally trivial} when
$ete=e$ for all $e\in E(\mathsf T)$ and $t\in\mathsf T$.
Every nonempty finite semigroup has an idempotent.
Consequently local triviality is preserved by
surjective finite-semigroup morphisms:
if $\pi:\mathsf T\to\mathsf S$ is surjective,
every idempotent of $\mathsf S$ lifts to one in
its preimage subsemigroup.

\begin{definition}[fixed-$s$ substitutability]\label{def:hsubst}
A language $L\subseteq\Sigma^*$ is \emph{$\sim_s$-substitutable} if for all
$x,y\in\Sigma^+$, equality $s(x)=s(y)$ together with
$\D_L(x)\cap\D_L(y)\ne\varnothing$ implies $\D_L(x)=\D_L(y)$.
\end{definition}
\begin{lemma}[canonical fresh-identity completion]\label{lem:fresh-identity}
Let $\mathsf S^\sharp=\mathsf S\cup\{\mathbf{1}_{\sharp}\}$
be the monoid obtained by adjoining a \emph{fresh}
identity $\mathbf{1}_{\sharp}$ to the finite semigroup $\mathsf S$,
even if $\mathsf S$ already has an identity.
There is a unique monoid extension
$\widehat s:\Sigma^*\twoheadrightarrow\mathsf S^\sharp$
of $s$, given by
\[
\widehat s(w)=
\begin{cases}
\mathbf{1}_{\sharp},&w=\lambda,\\
s(w),&w\in\Sigma^+.
\end{cases}
\]
It is surjective, and
$\mathsf S^\sharp\cong\Sigma^*/\ker\widehat s$.
Here $\ker\widehat s$ denotes the kernel congruence
\[
\ker\widehat s
=\{(x,y)\in\Sigma^*\times\Sigma^*:
\widehat s(x)=\widehat s(y)\}.
\]
In particular,
\[
\widehat s(x)=\widehat s(y)\quad\Longleftrightarrow\quad
(x=y=\lambda)\ \text{or}\
(x,y\in\Sigma^+\text{ and }s(x)=s(y)).
\]
Thus Definition~\ref{def:hsubst} is equivalent to
the all-fragment condition
\[
\forall x,y\in\Sigma^*,\quad
\bigl(\widehat s(x)=\widehat s(y)\land
\D_L(x)\cap\D_L(y)\ne\emptyset\bigr)
\Longrightarrow\D_L(x)=\D_L(y).
\]
Conversely, every finite-monoid typing
$h:\Sigma^*\to M$ induces the finite-semigroup typing
$s=h|_{\Sigma^+}:\Sigma^+\to h(\Sigma^+)$.
If substitutability relative to $h$ compares \emph{nonempty}
factors, as in Definition~\ref{def:hsubst}, it defines
exactly the same class and nonempty-factor reconstruction.
No equivalence is asserted for the different condition that
also compares $\lambda$ with nonempty factors of type $h(\lambda)$.
\end{lemma}
\begin{proof}
Adjoining the identity extends multiplication
and makes $\widehat s$ a monoid morphism.
Every element of $\mathsf S$ is attained by a
positive word, while $\lambda$ attains $\mathbf{1}_{\sharp}$,
so the map is surjective.  The displayed
kernel description follows because
$\mathbf{1}_{\sharp}\notin\mathsf S$.
Equal extended types compare either $\lambda$
with itself or nonempty words of equal $s$-type,
proving equivalence with Definition~\ref{def:hsubst}.
For the converse, restricting $h$ preserves
all nonempty equalities, which alone define
substitutability and the reconstruction rules.
\end{proof}

Throughout, subscripts of substitutability classes,
reconstruction operators, and typed-thickness parameters
denote semigroup morphisms on \emph{nonempty} words.
Write $s_0$ for the unique one-element semigroup typing,
on whichever alphabet is in use. When a monoid morphism
$h:\Sigma^*\to M$ is convenient for recognizing a regular
language or evaluating empty contexts, its restriction
$s=h|_{\Sigma^+}$ is the learning bias. The extension
$\widehat s$ handles nullable yields
(Lemma~\ref{lem:fresh-identity}); algebraic conditions
are tested on the positive image, not on its completion.

For $x\in\Sigma^*$ we call $\widehat s(x)$ its \emph{yield type};
for nonempty $x$ it equals $s(x)$.  As above, $\Ccf{s}$ denotes the context-free
$\sim_s$-substitutable languages, and we define
$\Clin{s}:=\{L\in\Ccf{s}:L\text{ is generated by some linear CFG}\}$.

\section{Fixed Recognizable Substitutability}
\label{sec:framework}

For fixed $s$, put
$\mathsf{RS}_s:=\{L\subseteq\Sigma^*:L\text{ is $\sim_s$-substitutable}\}$.
Then the fixed-$s$ slice is
$\Ccf{s}=\mathsf{RS}_s\cap\mathrm{CFL}$, while $\Clin{s}$ consists
of its linearly generated members.  For expressive comparisons only, let
$\mathsf{RS}:=\bigcup_s\mathsf{RS}_s$, where $s$ ranges over finite-semigroup
homomorphisms on $\Sigma^+$.  We write $\mathsf{RS}_s(\Gamma)$ and
$\mathsf{RS}(\Gamma)$ when the alphabet matters; all inclusions below are
alphabetwise.  The learning results always concern one fixed $s$.

\begin{proposition}\label{prop:regular-auto}
A language $L\subseteq\Sigma^*$ is regular iff
$L=\widehat s^{-1}(\mathsf{Acc})$ for some finite-semigroup
typing $s:\Sigma^+\to\mathsf S$ and some
$\mathsf{Acc}\subseteq\mathsf S^\sharp$.
Every language recognized in this way belongs to
$\mathsf{RS}_s$, so all regular languages belong to
$\mathsf{RS}$.
\end{proposition}
\begin{proof}
For regular $L$, restrict its syntactic-monoid
morphism~\cite[Chapter~III, Proposition~10.2]{eilenberg1974}
to $\Sigma^+$, replace the codomain by
its positive image $\mathsf S$, and extend by a
fresh identity.  The positive types still determine
membership for nonempty words; include $\mathbf{1}_{\sharp}$
among the accepting types exactly if $\lambda\in L$.
Conversely, $\widehat s$ recognizes a regular
language since $\mathsf S^\sharp$ is finite.
For nonempty $x,y$ with $s(x)=s(y)$,
$\widehat s(uxv)=\widehat s(uyv)$ for all contexts
$u,v\in\Sigma^*$, hence $\D_L(x)=\D_L(y)$.
\end{proof}

By Proposition~\ref{prop:regular-auto}, $\mathsf{RS}$ and
$\mathsf{RS}\cap\mathrm{CFL}$ contain every finite language and
$\Sigma^*$; Gold's theorem makes both unions nonlearnable from
positive data on nonempty alphabets.  Yoshinaka
\cite[Section~4]{yoshinaka2008} notes the analogous issue for
the union of fixed-window classes.  We learn only fixed-$s$ slices.

The fresh-identity completion reduces refinement and
product typing to ordinary monoid-kernel operations.
For finite-semigroup typings $s,s',t$ of $\Sigma^+$,
where each codomain is reduced to its positive image,
\[
\ker\widehat{s'}\subseteq\ker\widehat s
\ \Longleftrightarrow\
\ker s'\subseteq\ker s,\qquad
\ker\widehat{s\times t}
=\ker\widehat s\cap\ker\widehat t.
\]
Here $\widehat{s\times t}$ completes the \emph{image}
$(s\times t)(\Sigma^+)$ with a fresh identity.
In fact its image monoid is isomorphic to
$\operatorname{im}(\widehat s\times\widehat t)$,
a submonoid of $\mathsf S^\sharp\times\mathsf T^\sharp$.
Consequently $\mathsf{RS}_s\subseteq\mathsf{RS}_{s'}$
under the first inclusion, and
$\mathsf{RS}_s\cup\mathsf{RS}_t
\subseteq\mathsf{RS}_{s\times t}$.
A refinement may increase the typed grammar size
used in quantitative bounds.

For fixed $k,\ell$, Yoshinaka~\cite[Proposition~1]{yoshinaka2008}
showed that the class of all
\fwdlink{def:kl-subst}{fixed-window substitutable languages}
is closed under intersection of two members of the class and under
inverse images of nonerasing homomorphisms.
The same proposition shows that the class is not closed under
intersection with arbitrary regular languages or under arbitrary
inverse homomorphisms; the latter counterexample uses erasure.
Yoshinaka also notes that the context-free subclass need not be
closed under intersection.
Proposition~\ref{prop:finite-info-closure} instead records
finite-typing closure facts that permit regular filtering and
erasing inverse images by changing the typing.

\begin{proposition}[closure under finite information]\label{prop:finite-info-closure}
Let $s:\Sigma^+\to\mathsf S$ and
$t:\Sigma^+\to\mathsf T$ be finite-semigroup typings.
\begin{enumerate}[label=\textup{(\roman*)},nosep]
\item If $L\in\mathsf{RS}_s$ and $J\in\mathsf{RS}_t$, then
$L\cap J\in\mathsf{RS}_{s\times t}$.
\item Let $\mathsf{Acc}_t\subseteq\mathsf T^\sharp$ be arbitrary and put
$R_{\mathrm{reg}}:=\widehat t^{-1}(\mathsf{Acc}_t)$.
If $L\in\Ccf{s}$, then
$L\cap R_{\mathrm{reg}}\in\Ccf{s\times t}$.
\item Let $\varphi:\Gamma^*\to\Sigma^*$ be any
homomorphism, possibly erasing, and put
$r=(\widehat s\circ\varphi)|_{\Gamma^+}:
\Gamma^+\to\mathsf S^\sharp$.
If $L\in\mathsf{RS}_s$, then
$\varphi^{-1}(L)\in\mathsf{RS}_r$;
if moreover $L\in\Ccf{s}$, then
$\varphi^{-1}(L)\in\Ccf{r}$.\footnote{%
The fresh identity in $\mathsf S^\sharp$ replaces
a separate erasure flag.  For trivial $s$,
$L=a^+$, $\varphi(a)=a$ and $\varphi(b)=\lambda$,
the two nonempty inverse-image factors $a,b$
share a context but have different distributions.}
\end{enumerate}
\end{proposition}
\begin{proof}
For (i), nonempty factors of equal $(s\times t)$-type
with a common context in $L\cap J$ have equal
distributions in each language, and hence in
their intersection. Item (ii) follows from (i)
and closure of CFLs under regular intersection.
For (iii), $r$ is a semigroup morphism, since
$\widehat s\circ\varphi$ is a monoid morphism.
If two nonempty factors of equal $r$-type share
a context in $\varphi^{-1}(L)$, their images are
either both empty or both nonempty and of equal
$s$-type.  In the first case their inverse-image
distributions coincide.  In the second case,
substitutability of $L$ and the shared image
context give the same conclusion after pulling
back contexts.  The context-free assertion uses
closure of CFLs under inverse homomorphism.
\end{proof}

\paragraph{Typing adequacy.}
By Lemma~\ref{lem:fresh-identity},
$L\in\mathsf{RS}_s$ exactly when $\ker \widehat s$
distinguishes every pair of factors with intersecting but
unequal distributions.  Thus typing adequacy is a
finite-index congruence separation condition, preserved
under refinement and products at a possible quantitative
cost (Proposition~\ref{prop:typed-thickness-gap}).
The stronger condition $\ker \widehat s\subseteq\equiv_L$
is sufficient but forces $L$ to be regular.

\subsection{Classical special cases}

Clark--Eyraud's original Definition~5 includes all fragments
\cite{clark-eyraud2007}; we follow Clark~\cite{clark2013}
in excluding $\lambda$.  This distinction matters even for $L=a^+$:
$(a,\lambda)$ belongs to both $\D_L(\lambda)$ and $\D_L(a)$,
whereas $(\lambda,\lambda)$ belongs to $\D_L(a)$ but not to
$\D_L(\lambda)$.  Thus comparing $\lambda$ with nonempty factors
\emph{without type separation} would exclude $a^+$;
$\widehat s$ makes those factors incomparable in type.
Call $L$
\emph{Clark--Eyraud substitutable} when
$\D_L(x)\cap\D_L(y)\ne\emptyset$ implies
$\D_L(x)=\D_L(y)$ for all $x,y\in\Sigma^+$.  This is exactly the
trivial-semigroup case of Definition~\ref{def:hsubst}, and hence also
Yoshinaka's $(0,0)$ case.

For $k,\ell\ge0$, let $\pref_k(w)$ and $\suf_\ell(w)$ denote the
prefix and suffix of lengths $\min\{k,|w|\}$ and
$\min\{\ell,|w|\}$, respectively; in particular,
$\pref_0(w)=\suf_0(w)=\lambda$.  Renaming Yoshinaka's variables so that
$u_i,v_i$ denote outer contexts, \hypertarget{def:kl-subst}{}$L$ is \emph{$(k,\ell)$-substitutable}
if, for all $p\in\Sigma^k$, $q\in\Sigma^\ell$, and
$u_1,u_2,v_1,v_2,r_1,r_2\in\Sigma^*$ with $pr_1q,pr_2q\ne\lambda$,
\[
\begin{split}
&u_1pr_1qv_1\in L,\quad u_1pr_2qv_1\in L,\quad
u_2pr_1qv_2\in L\\
&\hspace{35mm}\Longrightarrow\quad u_2pr_2qv_2\in L.
\end{split}
\]

\begin{proposition}[finite-semigroup representation of fixed windows]
\label{prop:yl-special}
For every $k,\ell\ge0$ there is a surjective
finite-semigroup morphism $s_{k,\ell}:\Sigma^+\to W_{k,\ell}$
such that $L\subseteq\Sigma^*$ is $(k,\ell)$-substitutable
if and only if it is $\sim_{s_{k,\ell}}$-substitutable.
\end{proposition}
\begin{proof}
Put $m_0:=\max\{1,k+\ell\}$. For $w\in\Sigma^+$ define
\[
s_{k,\ell}(w):=
\begin{cases}
(\mathrm{short},w),&0<|w|<m_0,\\
(\mathrm{long},\pref_k(w),\suf_\ell(w)),&|w|\ge m_0.
\end{cases}
\]
The displayed equivalence on $\Sigma^+$ is a two-sided
congruence: equal short types represent the same word;
equal long types retain their length-$k$ prefix and
length-$\ell$ suffix after any common left and right
concatenation. Its finite quotient $W_{k,\ell}$ thus
inherits an associative product, and $s_{k,\ell}$
is a surjective semigroup morphism.

For nonempty $x,y$, equal types mean either $x=y$
or $x=pr_1q$, $y=pr_2q$ for common
$p\in\Sigma^k$, $q\in\Sigma^\ell$ and
$r_1,r_2\in\Sigma^*$. In the latter case a common
context and an arbitrary context of one factor
instantiate Yoshinaka's four-word condition;
symmetry gives equal distributions.
Conversely, the nonempty factors of any such
premise have equal $s_{k,\ell}$-types, so relative
substitutability supplies the fourth word.
\end{proof}

Yoshinaka~\cite[Example~4]{yoshinaka2015} already represented
$(k,\ell)$-substitutability by the sorted signature $\Sigma_{k,\ell}$.
Proposition~\ref{prop:yl-special} makes the inducing finite-semigroup
quotient explicit; no priority for the window-signature encoding is claimed.

\phantomsection\label{crit:kl}
\paragraph{Fixed-window counterexample criterion.}
If $x,y$ have length at least $\max\{1,k+\ell\}$, the same length-$k$
prefix and length-$\ell$ suffix, a common context, and unequal
distributions, then $L$ is not $(k,\ell)$-substitutable: the first
three conditions give equal $s_{k,\ell}$-types and a shared context, while
the last contradicts $\sim_{s_{k,\ell}}$-substitutability.

Thus the fixed $(k,\ell)$ class is exactly
$\mathsf{RS}_{s_{k,\ell}}$.  Section~\ref{sec:fixed-window-thick} gives
its polynomial thick-data bound.  For $k=\ell=0$,
$s_{0,0}$ maps all nonempty words to one type,
so it coincides with $s_0$ on $\Sigma^+$.  For an alphabet
$\Gamma$, set
\[
\mathsf{KL}(\Gamma):=
\bigcup_{k,\ell\ge0}
\{L\subseteq\Gamma^*:L\text{ is $(k,\ell)$-substitutable}\},
\]
and abbreviate $\mathsf{KL}:=\mathsf{KL}(\Sigma)$.

\subsection{Main theorem}

\begin{theorem}[fixed-$s$ learning theorem]\label{thm:main}
Fix an efficiently computable finite-semigroup homomorphism
$s:\Sigma^+\twoheadrightarrow\mathsf S$.  There are a set-driven reconstruction operator $\mathsf{Rec}_s$ and a
conservative sequential learner $\mathsf{Learn}_s$ such that:
\begin{enumerate}[label=\textup{(\roman*)},nosep]
\item $\mathsf{Rec}_s$ is sample-consistent and $\mathsf{Rec}_s(K)$ is
constructible in time polynomial in $\|K\|$;
\item every $L\in\Ccf{s}$ has a finite characteristic sample for
$\mathsf{Rec}_s$;
\item $\mathsf{Learn}_s$ identifies every nonempty $L\in\Ccf{s}$ in Gold's
sense, each update taking time polynomial in the data seen so far;
\item if the positive image semigroup $s(\Sigma^+)$ is a
homomorphic image of a subsemigroup of the direct product of
a finite locally trivial semigroup and a finite group,
then for every nonempty target the sample in~(ii) can be chosen polynomial in $|G_*|$
and $\tau_{G_*}+1$ for any reduced CFG $G_*$ representing that target;
this includes every fixed-window and every finite-group typing;
\item for arbitrary fixed $s$ and $L\in\Clin{s}$, the sample in~(ii) can
be chosen polynomial in $|G_*|$ for any linear CFG $G_*$ representing $L$.
\end{enumerate}
The polynomials in~(iv)--(v) may depend on the fixed $s$ and $\Sigma$,
but not on the chosen target grammar $G_*$. The quantitative mechanism behind items~(iv)--(v), together with its
limitation for arbitrary fixed typings, is developed in
Sections~\ref{sec:typed-thickness}--\ref{sec:linear}.
\end{theorem}

This is consistent with Gold's theorem: on a nonempty alphabet choose
distinct $x,y\in\Sigma^+$ with $s(x)=s(y)$, and choose
$z\in\Sigma^*$ with $|z|>\max\{|x|,|y|\}$.  The finite language $\{x,y,zx\}$
is not $\sim_s$-substitutable: $x,y$ share the empty context
but only $x$ admits $(z,\lambda)$.

\paragraph{Algebraic position of fixed windows.}
The locally trivial-semigroup identities underlying the
following criterion and the corresponding prefix--suffix
characterization of recognizable languages are classical
\cite[Chapter~2, Proposition~3.5 and Theorem~3.6]{pin1986};
see also the generalized definite language characterization
\cite[Proposition~5.18]{pin1997}.
We state the positive-image/kernel formulation explicitly for
arbitrary finite-semigroup typings and prove it below.
Its learning-specific application is Corollary~\ref{cor:li-thickness}.
These algebraic conditions must be tested on
$\mathsf S=s(\Sigma^+)$, \emph{not} on $\mathsf S^\sharp$.
Indeed, for nonempty $\mathsf S$, the new identity
$\mathbf{1}_{\sharp}$ and any $t\in \mathsf S$ satisfy
$\mathbf{1}_{\sharp} t\mathbf{1}_{\sharp}=t\ne\mathbf{1}_{\sharp}$;
thus $\mathsf S^\sharp$ itself is not locally trivial.

\begin{proposition}[locally trivial typings and fixed windows]
\label{prop:li-window}
Let $s:\Sigma^+\twoheadrightarrow\mathsf S$ be a
finite-semigroup typing on a nonempty alphabet and set
$n:=|\mathsf S|+1$. The following are equivalent:
\begin{enumerate}[label=\textup{(\roman*)},nosep]
\item $\mathsf S$ is locally trivial;
\item $\ker s_{k,\ell}\subseteq\ker s$ for some $k,\ell\ge0$.
\end{enumerate}
In (ii), $k=\ell=n$ suffices. Every window semigroup
$W_{k,\ell}=s_{k,\ell}(\Sigma^+)$ is locally trivial.
Consequently,
\[
\mathsf{KL}(\Sigma)=
\bigcup_{\substack{s:\Sigma^+\to\mathsf S\\
\mathsf S\text{ finite locally trivial}}}
\mathsf{RS}_s(\Sigma),
\]
also after intersecting both sides with $\mathrm{CFL}$.
\end{proposition}
\begin{proof}
A short-word class in $W_{k,\ell}$ cannot be idempotent:
squaring a positive word either changes its short
representative or reaches the long part. If $e=s_{k,\ell}(x)$
is a long type, then $xux$ retains the prefix and
suffix of $x$ for every $u\in\Sigma^+$.
Thus $ete=e$ for all $t\in W_{k,\ell}$.

For every finite semigroup $\mathsf S$, the classical identity
$\mathsf S^n=\mathsf S E(\mathsf S)\mathsf S$
holds for $n=|\mathsf S|+1$
\cite[Chapter~1, Proposition~1.12]{pin1986}.
For locally trivial $\mathsf S$, one also has
$etf=ef$ for all idempotents $e,f$ and $t\in\mathsf S$
\cite[Chapter~2, Proposition~3.5]{pin1986}.
For $p,q\in\Sigma^n$, write
$s(p)=\alpha e\beta$, $s(q)=\gamma f\delta$ with $e,f$ idempotent.
For every $r\in\Sigma^*$, including $\lambda$,
$\beta\widehat s(r)\gamma\in\mathsf S$, so
$s(prq)=\alpha e(\beta\widehat s(r)\gamma)f\delta
=\alpha ef\delta=s(pq)$.
Hence length-$n$ windows determine the type
of every word of length at least $2n$; shorter
words are separated by $s_{n,n}$. Conversely,
(ii) induces a surjective semigroup morphism
$W_{k,\ell}\to\mathsf S$, preserving local triviality.
The union identities follow by kernel refinement.
\end{proof}
\section{Reconstruction Operator and Sequential Learner}
\label{sec:algo}

\subsection{Finite-sample reconstruction}
For a finite sample $K\subseteq\Sigma^*$, let
\[
 \widehat V(K):=
 \{\NT{x}:x\in\Sigma^+,\ \exists u,v\in\Sigma^*\text{ with }uxv\in K\}.
\]
Thus there is one hypothesis nonterminal for each distinct nonempty factor
observed in the sample; the double brackets are part of the nonterminal name and do
not denote a quotient class or literal terminal symbols.  For observed nonempty factors $x,y$, write
\[
 x\approx_{K,s}y
 \quad\Longleftrightarrow\quad
 s(x)=s(y)\ \text{ and }\ \D_K(x)\cap\D_K(y)\ne\emptyset .
\]
The grammar $\mathsf{Rec}_s(K)$ has nonterminal set
$\widehat V(K)\cup\{\widehat S\}$, start symbol $\widehat S$, and the following
\emph{set} of productions whenever every displayed bracketed nonterminal belongs to
$\widehat V(K)$:
\[
\begin{array}{rll}
\textup{(B)}&\NT{xy}\to\NT{x}\NT{y},&x,y\in\Sigma^+,\\[1mm]
\textup{(U)}&\NT{x}\to\NT{y},&x\approx_{K,s}y,\\[1mm]
\textup{(L)}&\NT{a}\to a,&a\in\Sigma,\\[1mm]
\textup{(S)}&\widehat S\to\NT{w},&w\in K\cap\Sigma^+.
\end{array}
\]
If $\lambda\in K$, Rule~\textup{(S)} also contains
$\widehat S\to\lambda$.

For finite samples $K\subseteq K'$, every nonterminal and production
of $\mathsf{Rec}_s(K)$ also occurs in $\mathsf{Rec}_s(K')$.
Consequently,
$L(\mathsf{Rec}_s(K))\subseteq L(\mathsf{Rec}_s(K'))$.

The construction follows the substring-indexed grammar scheme of
Yoshinaka~\cite[Section~4, Algorithm~1]{yoshinaka2008}, also recalled
by Clark~\cite[Section~4, Algorithm~1]{clark2013}.
For the semigroup typing $s=s_{k,\ell}$, the nontrivial instances of Rule~\textup{(U)}
coincide with Yoshinaka's unary rules: distinct observed nonempty factors
are connected when they share a sample context and have equal
$s_{k,\ell}$-type; equivalently, both factors have length at least
$\max\{1,k+\ell\}$ and have the same length-$k$ prefix and
length-$\ell$ suffix.
The constructions agree up to redundant identity rules, the
representation of initial productions by a separate start symbol, and
the treatment of $\lambda$.  For general fixed $s$, the
prefix--suffix test is replaced by equality of $s$-types.

\begin{lemma}[sample consistency]\label{lem:sample-consistency}
For every finite sample $K$, $K\subseteq L(\mathsf{Rec}_s(K))$.
\end{lemma}
\begin{proof}
If $w=a_1\cdots a_m\in K\cap\Sigma^+$, Rule~\textup{(S)} enters $\NT{w}$.
Repeated use of Rule~\textup{(B)} along any binary factorization of $w$
reaches the one-letter factors, all of which are observed because they occur
inside the same sample word.  Rule~\textup{(L)} then emits
$a_1\cdots a_m$.  If $\lambda\in K$, the start rule emits it directly.
\end{proof}

\begin{theorem}[soundness]\label{thm:soundness}
If $K\subseteq L$ and $L$ is $\sim_s$-substitutable, then
$L(\mathsf{Rec}_s(K))\subseteq L$.
\end{theorem}

\begin{proof}
Induct on the height of a derivation tree rooted at $\NT{x}$ to prove
\begin{equation}\label{eq:soundness-invariant}
 \NT{x}\Rightarrow_{\mathsf{Rec}_s(K)}^*w
 \quad\Longrightarrow\quad
 x\equiv_L w\ \text{ and }\ s(w)=s(x).
\end{equation}
Every factor nonterminal is indexed by a nonempty observed word,
so the fragments compared by Rule~\textup{(U)} satisfy the nonempty
requirement of Definition~\ref{def:hsubst}.  Moreover,
Rules~\textup{(B)}--\textup{(L)} cannot derive $\lambda$ from a
factor nonterminal.  Rule~\textup{(L)} is immediate.  For Rule~\textup{(U)}, suppose
$\NT{x}\to\NT{y}$.  By definition there is a sample context $(u,v)$ with
$uxv,uyv\in K\subseteq L$ and $s(x)=s(y)$.  Fixed-$s$ substitutability gives
$x\equiv_L y$; the induction hypothesis gives $y\equiv_L w$ and
$s(w)=s(y)$, proving~\eqref{eq:soundness-invariant}.

For Rule~\textup{(B)}, write the derived word as $w_1w_2$ from
$\NT{xy}\to\NT{x}\NT{y}$.  The induction hypotheses give
$x\equiv_L w_1$, $y\equiv_L w_2$,
$s(w_1)=s(x)$, and $s(w_2)=s(y)$.
Because $\equiv_L$ is a monoid congruence,
$xy\equiv_L w_1w_2$, and the homomorphism law gives
$s(w_1w_2)=s(xy)$.

Finally, if a start rule $\widehat S\to\NT{x}$ is used, then $x\in K\subseteq L$,
so $(\lambda,\lambda)\in\D_L(x)$.  The invariant gives
$\D_L(w)=\D_L(x)$ and hence $w\in L$.  The direct $\lambda$ rule is present
only when $\lambda\in K\subseteq L$.
\end{proof}

\subsection{The sequential learner}

Recomputing $\mathsf{Rec}_s$ after every datum need not stabilize
syntactically: for $L_{\mathrm c}=\{a^ncb^n:n\ge0\}$, longer examples
introduce new factor nonterminals such as $\NT{a^ncb^n}$.
We therefore use the conservative wrapper of Clark--Eyraud's
Substitution Graph Learner (SGL, Algorithm~2) and Yoshinaka's
$(k,\ell)$-Substitutable Grammar Learner ($(k,\ell)$-SGL, Algorithm~1)~\cite{clark-eyraud2007,yoshinaka2008};
see also Eyraud--Heinz--Yoshinaka~\cite[Appendix]{eyraud-heinz-yoshinaka2016}
for the distinction between set-driven reconstruction and incremental learning.
Starting from the fixed empty grammar $H_0$, after reading $w_n$ put
\[
 H_n:=
 \begin{cases}
 H_{n-1},&w_n\in L(H_{n-1}),\\
 \mathsf{Rec}_s(K_n),&w_n\notin L(H_{n-1}),
 \end{cases}
 \qquad
 \mathsf{Learn}_s(\mathcal T[n]):=H_n.
\]
Every rebuild uses the whole accumulated sample $K_n$. Although
$\mathsf{Rec}_s$ is set-driven, $\mathsf{Learn}_s$ can stabilize on
different grammars for different presentation orders. Indeed, let $L\in\Ccf{s}$ be infinite and let
$\mathcal T$ be a text for $L$. Choose $w\in L$ longer than
every sample word observed at the learner's last rebuild.
The limiting grammar for $\mathcal T$ has no symbol $\NT{w}$,
whereas the text obtained by prepending $w$ rebuilds on a sample
containing $w$ and therefore retains $\NT{w}$ in its limiting
grammar. Both hypotheses generate $L$; only eventual syntactic
stabilization is asserted (Corollary~\ref{cor:ilt}).
See also Clark~\cite[Section~4, Example~1]{clark2013}.

\section{Finite Witnesses, Exact Reconstruction, and Complexity}
\label{sec:typing}

\subsection{Target refinement and finite witnesses}

For qualitative completeness, type an arbitrary target
CFG over $\mathsf S^\sharp$.  This refinement is used only in
the completeness proof to show that the already-defined hypothesis
grammar can simulate typed target derivations; it is not part of
the learning algorithm. The hypothesis grammar
$\mathsf{Rec}_s(K)$ remains unchanged, using nonempty observed
factors and a direct start rule for $\lambda$.

Let $G=(V,\Sigma,P,S_0)$ be a CFG for
a nonempty $L$.  Write a rule as
$\rho:A\to u_0B_1u_1\cdots B_ru_r$
with $u_i\in\Sigma^*$, $B_i\in V$, and $r\ge0$.
For each $(\mu_1,\ldots,\mu_r)\in(\mathsf S^\sharp)^r$,
set $\mu=\widehat s(u_0)\prod_{i=1}^r
(\mu_i\widehat s(u_i))$ and include
\[
 A_\mu\to u_0(B_1)_{\mu_1}u_1\cdots(B_r)_{\mu_r}u_r.
\]
This also covers terminal-only, empty, and unit rules.
Add a fresh start symbol $\widetilde S$ with
$\widetilde S\to(S_0)_\mu$ for each
$\mu\in\mathsf S^\sharp$, and trim to the reduced
grammar $\widetilde G$.  This refinement is finite
but can have exponential size; it is not part of
the learning algorithm.

Every retained typed symbol is reachable and productive.
Its yields have the indicated type; in particular,
$A_{\mathbf{1}_{\sharp}}$ yields only $\lambda$, while
$A_\mu$ with $\mu\ne\mathbf{1}_{\sharp}$ yields no empty word.
For each retained non-start symbol $X$, choose the
shortlex-minimal yield $\omega(X)$, possibly $\lambda$.
Typing ensures that every yield of $X$, in particular
$\omega(X)$, has the type assigned to $X$.
Surrounding contexts carry no type annotations.

\begin{proposition}[typed lifting and yield invariant]\label{prop:typed-core}
The typed refinement of any CFG $G$ satisfies
$L(\widetilde G)=L(G)$.  For every retained $A_\mu$,
including typed copies of $S_0$,
\[
L_{\widetilde G}(A_\mu)
 =L_G(A)\cap \widehat s^{-1}(\{\mu\}).
\]
In particular, $A_\mu\Rightarrow_{\widetilde G}^*w$
entails $\widehat s(w)=\mu$.
\end{proposition}
\begin{proof}
Erasing types gives a $G$-derivation; conversely,
annotating each node by its terminal-yield type
gives a typed derivation, including nullable children.
Successful typed derivations survive trimming.
For a retained $A_\mu$, graft any annotated
derivation of $w\in L_G(A)$ of type $\mu$
at a reachable occurrence of $A_\mu$.
This establishes the fiber and start-language equalities.
\end{proof}

For each retained non-start symbol $X$, choose a
\emph{canonical reaching context}
$\chi(X)=(u_X,v_X)$ minimizing $|u_X|+|v_X|$,
with shortlex tie breaks on $u_X$ and then $v_X$.
It exists because $\widetilde G$ was trimmed.  For each typed rule
$\rho:X\to u_0Y_1u_1\cdots Y_ru_r$
($X\ne\widetilde S$), let its
\emph{canonical right-hand-side yield} be
$t_\rho=u_0\omega(Y_1)u_1\cdots\omega(Y_r)u_r$.
Both $t_\rho$ and $\omega(X)$ belong to
$L_{\widetilde G}(X)$ and have the same type.
For nonempty-type $X$, these yields are nonempty and occur
in their shared canonical reaching context $\chi(X)$;
the corresponding witnesses therefore license the unary
Rule~\textup{(U)} from $\NT{\omega(X)}$ to $\NT{t_\rho}$.
For empty-type $X$, both yields are $\lambda$ and no factor
nonterminal or unary rule is needed. Thus typing is used for
this witness-based simulation, not claimed to be an
algorithm-independent necessity.
Define the finite set of \emph{canonical positive witnesses}
\begin{equation}\label{eq:W}
\begin{split}
\WitnessSet(\widetilde G):={}&
 \{u_X\omega(X)v_X:
     X\in\widetilde V\setminus\{\widetilde S\}\}\\
 &{}\cup\{u_Xt_\rho v_X:
     \rho=(X\to\alpha)\in\widetilde P,\
     X\ne\widetilde S\}.
\end{split}
\end{equation}
Every witness belongs to $L$.
For an empty-type rule, $t_\rho=\omega(X)=\lambda$,
so its witness is redundant.  If $\lambda\in L$,
the start copy $(S_0)_{\mathbf{1}_{\sharp}}$ has context
$(\lambda,\lambda)$ and yield $\lambda$; hence
$\lambda\in\WitnessSet(\widetilde G)$.

\subsection{Exact reconstruction and Gold convergence}

\begin{theorem}[completeness from the finite witnesses]\label{thm:complete}
Let $G$ be any CFG generating a nonempty language
$L$, and $\widetilde G$ its $\mathsf S^\sharp$-typed
refinement.  If $\WitnessSet(\widetilde G)\subseteq K$,
then $L\subseteq L(\mathsf{Rec}_s(K))$.
\end{theorem}
Neither substitutability nor $K\subseteq L$ is required.
\begin{proof}
For each retained nonempty-type $X=A_\mu$, put
$\widehat X:=\NT{\omega(X)}$, an observed nonempty factor
by the first witness family.  Induct on the height of a
typed derivation tree to show that
$X\Rightarrow_{\widetilde G}^*w$ implies
$\widehat X\Rightarrow_{\mathsf{Rec}_s(K)}^*w$.

Let the root rule be
$\rho:X\to u_0Y_1u_1\cdots Y_ru_r$, with child yields $w_i$.
Both $\omega(X)$ and
$t_\rho=u_0\omega(Y_1)u_1\cdots\omega(Y_r)u_r$
are nonempty, have the same $s$-type, and occur in
the shared context $(u_X,v_X)$ in $K$.
Thus Rule~\textup{(U)} gives $\widehat X\to\NT{t_\rho}$.

Delete the empty entries from
$(u_0,\omega(Y_1),u_1,\ldots,\omega(Y_r),u_r)$,
and denote the remaining entries in order by
$z_1,\ldots,z_\nu$.  Since $X$ has nonempty type,
$\nu\ge1$ and $t_\rho=z_1\cdots z_\nu$.
Each entry and every concatenation of consecutive
entries is an observed nonempty factor of
$u_Xt_\rho v_X\in K$.  Repeated Rule~\textup{(B)}
therefore gives
$\NT{t_\rho}\Rightarrow_{\mathsf{Rec}_s(K)}^*
\NT{z_1}\cdots\NT{z_\nu}$
(zero steps if $\nu=1$).
Terminal blocks are emitted by Rules~\textup{(B)}
and~\textup{(L)}.  For a positive-type child $Y_i$,
its corresponding entry is $z_j=\omega(Y_i)$,
so $\NT{z_j}=\widehat{Y_i}\Rightarrow^*w_i$
by induction.  Empty-type children have $w_i=\lambda$
and are omitted.  Hence
$\widehat X\Rightarrow^*
u_0w_1u_1\cdots w_ru_r=w$.

Every nonempty target word $w$ has a typed start
copy $(S_0)_{\widehat s(w)}$ with canonical context
$(\lambda,\lambda)$.  Its canonical yield belongs
to $K$, so start Rule~\textup{(S)} and the induction
give a derivation of $w$.  If $\lambda\in L$,
its start-copy witness puts $\lambda\in K$,
and Rule~\textup{(S)} emits it directly.
\end{proof}

\begin{theorem}[exact finite-sample reconstruction]
\label{thm:reconstruction-fixed-s}
Let $L\in\Ccf{s}$ be nonempty and let $G$ be
any CFG for $L$, with
$\mathsf S^\sharp$-typed refinement $\widetilde G$.
If $\WitnessSet(\widetilde G)\subseteq K\subseteq L$,
then $L(\mathsf{Rec}_s(K))=L$.
\end{theorem}
\begin{proof}
Apply Theorem~\ref{thm:complete} to $G$
and Theorem~\ref{thm:soundness} to $K\subseteq L$.
\end{proof}

Thus $\WitnessSet(\widetilde G)$ is characteristic
for every nonempty target, including $\{\lambda\}$.
For $L=\emptyset$, the empty sample gives
$L(\mathsf{Rec}_s(\emptyset))=\emptyset$.
This proves Theorem~\ref{thm:main}(ii).

\begin{corollary}[Gold identification]\label{cor:ilt}
For a fixed surjective semigroup homomorphism
$s:\Sigma^+\to\mathsf S$ to a finite semigroup,
the learner $\mathsf{Learn}_s$ identifies every
nonempty language in $\Ccf{s}$ from positive data.
More precisely, for any CFG $G$ representing
the target and its general typed refinement
$\widetilde G$, once
$\WitnessSet(\widetilde G)\subseteq K_{n_0}$,
among stages $n\ge n_0$ at most one hypothesis
change can occur.
\end{corollary}

\begin{proof}
Let $\mathcal T$ be a text for a nonempty $L$, and choose $n_0$ with
$\WitnessSet(\widetilde G)\subseteq K_{n_0}$.  Since
$w_1\notin L(H_0)=\emptyset$, the learner rebuilds at stage $1$, so for
$n\ge1$ the hypothesis $H_n$ is $\mathsf{Rec}_s(K_m)$ for the most recent
rebuild stage $m\le n$.  Sample consistency of that rebuild and the rebuild
rule for the later data give $K_n\subseteq L(H_n)$, and soundness gives
$L(H_n)\subseteq L$, as $K_m\subseteq L$.  If a rebuild occurs at some
$n\ge n_0$, Theorem~\ref{thm:reconstruction-fixed-s} gives $L(H_n)=L$, and
no later datum triggers a change.  Otherwise $H_n=H_{n_0}$ for all
$n\ge n_0$, so every word of $\mathcal T$ lies in some
$K_n\subseteq L(H_n)=L(H_{n_0})$ with $n\ge n_0$; since every word of $L$
occurs in $\mathcal T$, $L(H_{n_0})=L$.  In both cases at most one change occurs
after stage $n_0$.
\end{proof}

\subsection{Complexity of the reconstruction step}
\label{sec:complexity}

\begin{theorem}[polynomial-time reconstruction]\label{thm:poly-build}
For fixed $s$, $\mathsf{Rec}_s(K)$ is constructible in time polynomial in
$\|K\|$.
\end{theorem}
\begin{proof}
Put $n_K:=\|K\|$ and let
$\mathcal O_K:=
\{(u,x,v)\in\Sigma^*\times\Sigma^+\times\Sigma^*
:uxv\in K\}$
be the set of observed nonempty factor occurrences.
Since $|\mathcal O_K|=O(n_K^2)$, there are at most
$O(n_K^2)$ factor nonterminals.
Extending factors one symbol at a time from each starting
position of each sample word computes and caches all their
$s$-values using $O(n_K^2)$ multiplications in the fixed semigroup.

Rule~\textup{(B)} has at most $O(n_K^3)$ candidates,
obtained by choosing an observed factor occurrence and a split point.
For Rule~\textup{(U)}, for each $(u,x,v)\in\mathcal O_K$
and each $w\in K$, check whether there is a nonempty word $y$
with $w=uyv$. Such a $y$, if it exists, is unique;
if $s(x)=s(y)$, emit the rule $\NT{x}\to\NT{y}$.
This enumerates all Rule~\textup{(U)} candidates, at most
$|\mathcal O_K||K|=O(n_K^3)$ of them.

Lexical and start rules are lower-order.  Checking, encoding and
inserting a candidate take $O(n_K)$ time, and a trie of the explicit rule
encodings removes duplicates, so the output is a rule set independent of
the enumeration order.  Thus $\mathsf{Rec}_s(K)$ is constructible in
$O(n_K^4)$ time.
\end{proof}

Each update of $\mathsf{Learn}_s$ is also polynomial in the data seen so far,
since CFG membership is polynomial and hypotheses and rebuilds are
polynomial by Theorem~\ref{thm:poly-build}.
Together with sample consistency (Section~\ref{sec:algo}),
Theorem~\ref{thm:reconstruction-fixed-s} and Corollary~\ref{cor:ilt}, this
proves Theorem~\ref{thm:main}(i)--(iii); items~(iv)--(v) are proved in
Sections~\ref{sec:fixed-window-thick}--\ref{sec:linear}.  Grammar size alone
cannot bound witness lengths: $S_0\to D_n$, $D_0\to a$,
$D_i\to D_{i-1}D_{i-1}$ has size $O(n)$ but generates
$\{a^{2^n}\}$, substitutable for every $s$
\cite[Example~2]{clark-eyraud2007},
\cite[Example~2.3]{eyraud-heinz-yoshinaka2016}.

\section{Typed Thickness for General Fixed Typings}
\label{sec:typed-thickness}

\begin{definition}[binary--unit--nullable normal form]
A CFG is in \emph{binary--unit--nullable normal form} (BUNF)
if every production is $A\to BC$, $A\to B$,
$A\to a$, or $A\to\lambda$.
\end{definition}
Terminal isolation and binarization give a polynomial-size
equivalent BUNF grammar; Proposition~\ref{prop:thick-bunf-normal}
also controls ordinary thickness.  Here BUNF bounds
typed-rule counts and derivation-tree branching,
without restricting Section~\ref{sec:typing}'s learning results.
Let $G=(V,\Sigma,P,S_0)$ be a reduced BUNF grammar,
$\widetilde G$ its refinement, and put
$N:=|V|$ and
$N_t:=|\widetilde V\setminus\{\widetilde S\}|$.
Define its \emph{typed thickness} by
\[
\tau_s^{\mathrm{typ}}(G):=
\max\Bigl(\{\,|\omega(X)|:
X\in\widetilde V\setminus\{\widetilde S\}\,\}\cup\{1\}\Bigr).
\]
It is a presentation-dependent parameter, not an
additional assumption on the learner.
The typed thickness includes copies of the original start symbol,
even if it occurs on a right-hand side or derives $\lambda$.
In particular, every nonempty $L=L(G)$ satisfies
\begin{equation}\tag{$\spadesuit$}\label{eq:typed-fiber-min}
 \tau_s^{\mathrm{typ}}(G)\ge
 \max_{\mu\in\widehat s(L)}
 \min\{\,|w|:w\in L,\ \widehat s(w)=\mu\,\}.
\end{equation}
Indeed, for each $\mu\in\widehat s(L)$, the typed start copy
$(S_0)_\mu$ survives trimming, and
Proposition~\ref{prop:typed-core} identifies its shortest yield length
with the corresponding minimum on the right.

\begin{lemma}[typed-thickness witness bound]
\label{lem:typed-thickness-bound}
Let $G$ and $\widetilde G$ be as above, and put
$\tau:=\tau_s^{\mathrm{typ}}(G)$.
Every non-start typed symbol $X$ satisfies
$|u_X|+|v_X|\le(N_t-1)\tau$, and each
$w\in\WitnessSet(\widetilde G)$ satisfies
$|w|\le(N_t+1)\tau$.
Let $P_{\mathrm{ter}}$, $P_{\mathrm{unit}}$ and
$P_{\mathrm{bin}}$ be the terminal, unit and binary
productions of $G$.  Then
\[
\|\WitnessSet(\widetilde G)\|
\le\bigl(N|\mathsf S^\sharp|+|P_{\mathrm{ter}}|
 +|P_{\mathrm{unit}}||\mathsf S^\sharp|
 +|P_{\mathrm{bin}}||\mathsf S^\sharp|^2\bigr)
 \bigl((N|\mathsf S^\sharp|+1)\tau+1\bigr).
\]
For fixed $s$, this is polynomial in $|G|$ and
$\tau_s^{\mathrm{typ}}(G)$.
\end{lemma}
\begin{proof}
In the dependency graph of retained typed symbols
include unit and binary dependencies.  Each symbol
is reachable from a typed copy of $S_0$; a shortest
path has at most $N_t-1$ edges.  Expand off-path
siblings of binary rules by canonical yields,
each of length at most $\tau$.  Unit steps contribute
no sibling.  Hence the canonical reaching context has
length at most $(N_t-1)\tau$.
Each BUNF rule adds at most two canonical yields or
one terminal.  Empty-type rule witnesses duplicate
canonical-yield witnesses and need no extra count.
There are at most $N|\mathsf S^\sharp|$ typed symbols,
$|P_{\mathrm{ter}}|$ terminal typed rules,
$|P_{\mathrm{unit}}||\mathsf S^\sharp|$ unit typed rules, and
$|P_{\mathrm{bin}}||\mathsf S^\sharp|^2$ binary typed rules.
Summing $|w|+1$ over the witness set proves the claim.
\end{proof}

\begin{corollary}[characteristic data via BUNF normalization]
\label{cor:typed-thickness-data}
Fix $s$.  Let $G_*$ be any reduced CFG for a nonempty
$L(G_*)\in\Ccf{s}$, and let $G$ be an equivalent reduced
BUNF normalization of size polynomial in $|G_*|$.
(If $G_*$ is already BUNF, take $G=G_*$.)
Then $\WitnessSet(\widetilde G)$ is a characteristic sample
for $\mathsf{Rec}_s$ whose encoded size is polynomial in
$|G_*|$ and $\tau_s^{\mathrm{typ}}(G)$.
\end{corollary}

\begin{proof}
Applying Theorem~\ref{thm:reconstruction-fixed-s} to $G$
shows that $\WitnessSet(\widetilde G)$ is characteristic
for $L(G)=L(G_*)$.
Apply Lemma~\ref{lem:typed-thickness-bound} to its chosen
BUNF normalization $G$, using that $|G|$ is polynomial
in $|G_*|$.
\end{proof}

The typed-thickness parameter in Corollary~\ref{cor:typed-thickness-data}
belongs to the chosen BUNF normalization, not directly to $G_*$.
For arbitrary fixed $s$, this does not imply a polynomial bound
in the original grammar size and ordinary thickness; indeed,
ordinary thickness cannot in general replace typed thickness
even for the reconstruction operator.

\paragraph{A bracketed doubling family.}
Let $\Gamma_{\mathrm{br}}=\{\mathtt a,\lfloor,\rfloor,\lceil,\rceil\}$.
Let $U_1=\{1,\zeta\}$ be the two-element semilattice
with identity $1$ and absorbing element $\zeta$:
$1\zeta=\zeta1=\zeta^2=\zeta$.
Define the surjective semigroup morphism
$s_{\mathrm U}:\Gamma_{\mathrm{br}}^+\to U_1$ by sending
$\mathtt a,\lfloor,\rfloor$ to $1$ and
$\lceil,\rceil$ to $\zeta$.
Thus $s_{\mathrm U}(w)=1$ exactly for positive
words without ceiling brackets, whereas
$\widehat{s}_{\mathrm U}(\lambda)=\mathbf{1}_{\sharp}\ne1$.

For $n\ge1$, define the reduced CFG $G_n$ by
\[
 S_0\to E_n,\qquad E_0\to\mathtt a,\qquad
 E_i\to\lfloor{}E_{i-1}E_{i-1}\rfloor
       \mid\lceil{}E_{i-1}\rceil
 \quad(1\le i\le n),
\]
and put $L_n=L(G_n)$. Then $|G_n|=O(n)$ and
$\tau_{G_n}=2n+1$.
Write $w^*_0=\mathtt a$ and
$w^*_i=\lfloor{}w^*_{i-1}w^*_{i-1}\rfloor$.
The only word of $L_n$ without ceiling brackets is $w^*_n$, of length
$|w^*_n|=3\cdot2^n-2$: without unary ceiling-bracket nodes the
derivation is the unique complete binary tree of height $n$.

\begin{proposition}[refinement can force exponential characteristic data]
\label{prop:typed-thickness-gap}
For every $n\ge1$ the following hold.
\begin{enumerate}[label=\textup{(\roman*)},nosep]
\item $L_n$ is Clark--Eyraud substitutable, and hence
$L_n\in\Ccf{s_0}\cap\Ccf{s_{\mathrm U}}$.
\item $L_n$ has a characteristic sample for $\mathsf{Rec}_{s_0}$
of encoded size polynomial in $n$.
\item Every finite $K\subseteq L_n$ such that
$w^*_n\in L(\mathsf{Rec}_{s_{\mathrm U}}(K))$ contains $w^*_n$.
Consequently, every characteristic sample $\mathcal F$ of $L_n$ for
$\mathsf{Rec}_{s_{\mathrm U}}$ satisfies
$\|\mathcal F\|\ge 3\cdot 2^n-1$.
\item Put $L^-_n:=L_n\setminus\{w_n^*\}$. Then
$L^-_n\in\Ccf{s_{\mathrm U}}$. For any set-driven reconstruction
operator $\mathsf{Op}$ with characteristic samples for both $L_n$
and $L^-_n$, every characteristic sample $\mathcal F$ of $L_n$ for
$\mathsf{Op}$ contains $w_n^*$ and hence satisfies
$\|\mathcal F\|\ge3\cdot2^n-1$. In particular, this holds for every
set-driven operator having characteristic samples for the entire
fixed-$s_{\mathrm U}$ class.
\end{enumerate}
In particular, for the fixed typing $s_{\mathrm U}$ no bound on
characteristic-data size for $\mathsf{Rec}_{s_{\mathrm U}}$
polynomial in source grammar size and ordinary thickness holds over
all reduced CFG presentations of targets in $\Ccf{s_{\mathrm U}}$.
\end{proposition}

\begin{proof}
The substitutability assertion (i) is proved in
Appendix~\ref{app:bracketed-substitutability}.
For (ii), an explicit sample suffices. Set $u_0=\mathtt a$,
$u_i=\lceil{}u_{i-1}\rceil$, and
\[
\mathcal F_n=\{u_n\}\cup
\{\lceil^{n-i}\lfloor u_{i-1}u_{i-1}\rfloor
\rceil^{n-i}:1\le i\le n\}.
\]
For each $i$, the two sample words provide a common context for
$u_i$ and $\lfloor u_{i-1}u_{i-1}\rfloor$;
Rules~\textup{(U)} and \textup{(B)} therefore recover both
$E_i$ alternatives, so
$L_n\subseteq L(\mathsf{Rec}_{s_0}(\mathcal F_n))$.
For any $\mathcal F_n\subseteq K\subseteq L_n$, monotonicity
applied to $\mathcal F_n\subseteq K$ gives
$L_n\subseteq L(\mathsf{Rec}_{s_0}(K))$, while soundness
(Theorem~\ref{thm:soundness}) applied to $K\subseteq L_n$ gives
$L(\mathsf{Rec}_{s_0}(K))\subseteq L_n$.
Thus $\mathcal F_n$ is characteristic, with
$|\mathcal F_n|=n+1$ and $\|\mathcal F_n\|=3n^2+4n+2$.

For (iii)--(iv), the deleted-word target is the type-fiber
filter
$L_n^-=L_n\cap\widehat{s}_{\mathrm U}^{-1}(\{\zeta\})$.
It belongs to $\Ccf{s_{\mathrm U}}$: equal
$s_{\mathrm U}$-types of nonempty factors remain
equal $\widehat{s}_{\mathrm U}$-types inside any
common outer context. Thus intersection with
a union of $\widehat{s}_{\mathrm U}$-fibers
preserves relative substitutability, and the
intersection is context-free by regular filtering.

For (iii), if $K\subseteq L_n$ omits $w_n^*$, then
$K\subseteq L_n^-$. Soundness (Theorem~\ref{thm:soundness})
gives $L(\mathsf{Rec}_{s_{\mathrm U}}(K))\subseteq L_n^-$,
so $w_n^*$ cannot be generated. Hence every characteristic
sample $\mathcal F$ contains $w_n^*$ and satisfies
$\|\mathcal F\|\ge |w_n^*|+1=3\cdot2^n-1$.
Together with $|G_n|,\tau_{G_n}=O(n)$, this proves
the final assertion.

For (iv), let $\mathcal F$ be characteristic for $L_n$ under
$\mathsf{Op}$ and suppose $w_n^*\notin\mathcal F$.
Take a characteristic sample $\mathcal F'$ for $L_n^-$.
Then $\mathcal F\cup\mathcal F'\subseteq L_n^-\subsetneq L_n$,
so the same output $\mathsf{Op}(\mathcal F\cup\mathcal F')$
would have to generate both languages, a contradiction.

For $n\ge2$, $L_n^-\notin\Ccf{s_0}$:
$u_{n-1}$ and $w_{n-1}^*$ share the context
$(\lfloor,u_{n-1}\rfloor)$ in $L_n^-$, but only the former
admits $(\lfloor w_{n-1}^*,\rfloor)$.
Thus (iv) does not conflict with the $s_0$ upper bound in (ii).
\end{proof}

\begin{remark}[finite targets and linear grammar size]
Although each $L_n$ is finite and hence linear as a language,
the displayed $O(n)$-size grammar $G_n$ is not linear.
For a finite language generated by a linear CFG, a productive
derivation with a repeated nonterminal on its unique spine
cannot produce a nonempty terminal string around the cycle:
iterating such a cycle would generate infinitely many words.
Cycles producing only the empty word can be omitted.
Consequently a derivation can be chosen without repeated
nonterminals, and its terminal word length is bounded by
the total number of terminal occurrences in the grammar's
production listing. Since $|w_n^*|=3\cdot2^n-2$, every
linear CFG generating $L_n$ has exponentially many terminal
occurrences in its productions. This is consistent with
Theorem~\ref{thm:linear-poly}.
\end{remark}

The contradiction in part~(iv) follows the classical distinguishing-pair
principle using a word in the symmetric difference of two targets;
compare~\cite[Theorem~15.1.4]{higuera2010}.  The fixed-$s$
distinguishing pair $L_n,L_n^-$ and its exponential witness length
are specific to the present construction.

\begin{corollary}[presentation-independent typed-thickness gap]
\label{cor:typed-gap-intrinsic}
Every reduced BUNF grammar $G=(V,\Gamma_{\mathrm{br}},P,S_0)$
for $L_n$ satisfies
\[
 \boxed{\tau_{s_{\mathrm U}}^{\mathrm{typ}}(G)\ge3\cdot2^n-2.}
\]
This holds regardless of grammar size. Conversely, $L_n$
has a reduced BUNF presentation with both size and ordinary
thickness polynomial in $n$.
\end{corollary}
\begin{proof}
The positive fiber of type $1$ is
$L_n\cap\widehat{s}_{\mathrm U}^{-1}(\{1\})=\{w_n^*\}$,
with $1\ne\mathbf{1}_{\sharp}$.
The bound~\eqref{eq:typed-fiber-min} therefore gives
$\tau_{s_{\mathrm U}}^{\mathrm{typ}}(G)\ge
|w_n^*|=3\cdot2^n-2$ for every such $G$.
For the ordinary-thickness upper construction,
apply Proposition~\ref{prop:thick-bunf-normal} to $G_n$.
\end{proof}

\section{Polynomial Thick Data Beyond Fixed Windows}
\label{sec:fixed-window-thick}

Yoshinaka~\cite[Theorem~1]{yoshinaka2008} already established
polynomial thick-data learnability for fixed $(k,\ell)$-substitutable
context-free languages using $(k,\ell)$-Greibach normal form
and characteristic-set bounds~\cite[Lemmas~2 and~7]{yoshinaka2008}.
Here we instead bound the shortest yields of the typed refinement used by
$\mathsf{Rec}_s$.  One argument covers both fixed windows and
typings that additionally track finite-group values.
The corresponding batch bounds establish
identification in polynomial time and thick data (IPTtD),
in the set-driven sense of~\cite{yoshinaka2008} and
\cite[Definition~2.12]{eyraud-heinz-yoshinaka2016}.
They do not assert the stronger incremental stabilization requirement
of \cite[Definition~2.5]{eyraud-heinz-yoshinaka2016}.
We claim no improvement in Yoshinaka's polynomial degree.
For general fixed $s$, ordinary thickness need not give a polynomial
characteristic-data bound for $\mathsf{Rec}_s$
(Proposition~\ref{prop:typed-thickness-gap}).

Fix $k,\ell\ge0$ and a morphism $g:\Sigma^*\to\mathsf H$
into a finite group. In products with $s_{k,\ell}$,
the group component is restricted to $\Sigma^+$.
\begin{lemma}[window--group typed-yield bound]
\label{lem:window-typed-yield}
Let $G=(V,\Sigma,P,S_0)$ be a reduced BUNF grammar,
$N:=|V|\ge1$, $m:=k+\ell$ and $q:=|\mathsf H|$.
If $\ker\bigl(s_{k,\ell}\times(g|_{\Sigma^+})\bigr)
 \subseteq\ker s$, then
\[
 \tau_s^{\mathrm{typ}}(G)
 \le m+1+[1+2(m+q)(m+2)Nq](\tau_G+1).
\]
This includes nullable and unit productions, including
productions whose right-hand sides contain $S_0$.
\end{lemma}
\begin{proof}
See Appendix~\ref{app:fixed-window-bounds}, where a
nonemptiness anchor and derivation-size tie break
handle the additional BUNF cases.
\end{proof}

\begin{theorem}[polynomial thick data under window--group refinement]
\label{thm:window-thick}
Fix $s,\Sigma$ and suppose
$\ker\bigl(s_{k,\ell}\times(g|_{\Sigma^+})\bigr)
 \subseteq\ker s$
for some fixed $k,\ell$ and finite-group morphism $g$.
Every nonempty $L\in\Ccf{s}$ represented by a
reduced BUNF grammar $G$ has a characteristic sample
$\WitnessSet(\widetilde G)$ for $\mathsf{Rec}_s$
of encoded size polynomial in $|G|$ and $\tau_G+1$.
\end{theorem}
\begin{proof}
Combine Lemma~\ref{lem:window-typed-yield} and
Corollary~\ref{cor:typed-thickness-data}.
\end{proof}

Theorem~\ref{thm:window-thick} includes fixed windows
when $\mathsf H$ is trivial. For arbitrary CFG presentations
we need only terminal isolation and binarization.

\begin{proposition}[polynomial thickness-preserving BUNF normalization]
\label{prop:thick-bunf-normal}
There are fixed polynomials $q_{\mathrm{size}}$ and
$q_{\mathrm{thick}}$ such that every reduced CFG $G_*$
generating a nonempty language can be converted in
polynomial time into an equivalent reduced BUNF grammar $G$
with
$|G|\le q_{\mathrm{size}}(|G_*|)$ and
$\tau_G\le q_{\mathrm{thick}}(|G_*|,\tau_{G_*}+1)$.
The construction retains nullable and unit productions.
\end{proposition}
See Appendix~\ref{app:thick-bunf}.

Recall that a semigroup $\mathsf S$ \emph{divides} a semigroup $\mathsf Q$
if $\mathsf S$ is a homomorphic image of a subsemigroup of $\mathsf Q$.

\begin{lemma}[locally trivial--group divisors and window refinements]
\label{lem:li-group-factor}
Let $s:\Sigma^+\twoheadrightarrow\mathsf S$
be a finite-semigroup typing on a nonempty alphabet.
The following are equivalent:
\begin{enumerate}[label=\textup{(\roman*)},nosep]
\item $\mathsf S$ divides a product $\mathsf T\times\mathsf H$,
where $\mathsf T$ is a finite locally trivial semigroup
and $\mathsf H$ is a finite group;
\item for some $k,\ell\ge0$ and a monoid morphism
$g:\Sigma^*\to\mathsf H$ into a finite group $\mathsf H$,
\[
\ker\bigl(s_{k,\ell}\times(g|_{\Sigma^+})\bigr)
\subseteq\ker s.
\]
\end{enumerate}
\end{lemma}
\begin{proof}
For (i), choose a subsemigroup
$\mathsf U\subseteq\mathsf T\times\mathsf H$ and a surjective
morphism $\pi:\mathsf U\to\mathsf S$. Lift the letter images
$s(a)$ to $\mathsf U$, obtaining $\varphi:\Sigma^+\to\mathsf U$
with $\pi\varphi=s$. The factor projections are
$\sigma:\Sigma^+\to\mathsf T$ and
$g_+:\Sigma^+\to\mathsf H$. The latter extends
to $g:\Sigma^*\to\mathsf H$ with
$g(\lambda)=1_{\mathsf H}$. Since
$\sigma(\Sigma^+)$ is locally trivial,
Proposition~\ref{prop:li-window} supplies
$k,\ell$ with $\ker s_{k,\ell}\subseteq\ker\sigma$.
Together with $g_+$ this implies (ii).

Conversely (ii) induces a surjective semigroup morphism
\[
(s_{k,\ell}\times g)(\Sigma^+)\longrightarrow\mathsf S,
\qquad(s_{k,\ell}(w),g(w))\longmapsto s(w).
\]
The domain is a subsemigroup of
$W_{k,\ell}\times\mathsf H$, with $W_{k,\ell}$
locally trivial by Proposition~\ref{prop:li-window};
hence (i).
\end{proof}

\begin{corollary}[ordinary-thickness bounds for locally trivial--group typings]
\label{cor:li-thickness}
Fix $s:\Sigma^+\twoheadrightarrow\mathsf S$. Suppose $\mathsf S$ is a homomorphic
image of a subsemigroup of the direct product of a finite locally
trivial semigroup and a finite group.
Then every nonempty $L\in\Ccf{s}$ represented by a reduced CFG $G_*$
has characteristic data for $\mathsf{Rec}_s$ of encoded size
polynomial in $|G_*|$ and $\tau_{G_*}+1$.
This includes all locally trivial positive-image typings and
all finite-group typings.
\end{corollary}
\begin{proof}
By Lemma~\ref{lem:li-group-factor}, Theorem~\ref{thm:window-thick}
applies to the BUNF grammar given by
Proposition~\ref{prop:thick-bunf-normal}; compose the polynomials.

\end{proof}

Let $\mathbf{LI}$ and $\mathbf{Grp}$ be the pseudovarieties of finite locally
trivial semigroups and of finite groups.  As both are closed under finite
direct products, their join $\mathbf{LI}\vee\mathbf{Grp}$ consists exactly of
the divisors (homomorphic images of subsemigroups) of products $\mathsf{T}\times \mathsf{H}$
with $\mathsf{T}\in\mathbf{LI}$ and $\mathsf{H}\in\mathbf{Grp}$; such joins were studied by
Costa~\cite{costa2002li,costa2002lig}.  The hypothesis of
Corollary~\ref{cor:li-thickness} thus reads
$s(\Sigma^+)\in\mathbf{LI}\vee\mathbf{Grp}$.

\begin{lemma}[the two-element semilattice lies outside the window--group join]
\label{lem:U1-outside}
$U_1\notin\mathbf{LI}\vee\mathbf{Grp}$.
\end{lemma}

\begin{proof}
Suppose there were a subsemigroup $\mathsf{V}\subseteq \mathsf{T}\times \mathsf{H}$ and a
surjective morphism $\pi:\mathsf{V}\to U_1$, with
$\mathsf{T}\in\mathbf{LI}$ and $\mathsf{H}$ a finite group.
Choose idempotents $e\in\pi^{-1}(1)$ and $f\in\pi^{-1}(\zeta)$
(Section~\ref{sec:fixed-typing}).
Every idempotent of $\mathsf{T}\times \mathsf{H}$ has group component $1_{\mathsf{H}}$,
so the $\mathsf{T}$-components of $e,f$ are idempotents.
Local triviality of $\mathsf{T}$ gives $efe=e$.
Applying $\pi$ would yield
$1\cdot\zeta\cdot1=1$, contradicting $\zeta\ne1$.
\end{proof}

\begin{remark}
The positive image of $s_{\mathrm U}$ is $U_1$, which lies outside
$\mathbf{LI}\vee\mathbf{Grp}$ by Lemma~\ref{lem:U1-outside}.
Proposition~\ref{prop:typed-thickness-gap} shows that the
ordinary-thickness characteristic-data bound in
Corollary~\ref{cor:li-thickness} can fail for
$\mathsf{Rec}_s$ when its algebraic hypothesis is removed.
This does not establish that the hypothesis is necessary
for every fixed typing. Proposition~\ref{prop:typed-thickness-gap}(iv)
does give a reconstruction-operator-independent lower bound
on characteristic-sample size for the distinguishing target
pair, but not a lower bound on the running time or convergence
stage of arbitrary Gold learners.
\end{remark}

These polynomial bounds may depend on the fixed typing and alphabet;
no uniform bound for variable windows or groups is claimed.

\section{The Linear Subclass}
\label{sec:linear}

Linear languages are exactly the languages accepted by one-turn pushdown
automata~\cite{ginsburg-spanier1966,autebert1997}.  The normalization below
gives the grammatical counterpart: every successful derivation has one
continuing nonterminal path, which we call its \emph{spine}.

\begin{proposition}[linear-spine BUNF normalization]\label{prop:linear-normal}
There is a fixed polynomial $q_{\mathrm{lin}}$ such that every linear CFG
$G$ generating a nonempty language can be converted in polynomial time into
an equivalent reduced BUNF grammar $G_0=(V_0,\Sigma,P_0,S_0)$ with
$|G_0|\le q_{\mathrm{lin}}(|G|)$ in which every binary production has
exactly one fresh terminal-wrapper child $W_a$ with $W_a\to a$ and exactly
one other nonterminal child.
\end{proposition}

The construction is given in Appendix~\ref{app:linear-normalization}.  Let
$\widetilde G_0$ be its typed refinement and
$N_t:=|\widetilde V_0\setminus\{\widetilde S\}|\le|V_0|\,|\mathsf S^\sharp|$.

\begin{lemma}[short canonical yields]\label{lem:linear-short}
Every typed non-start symbol $X$ of $\widetilde G_0$ satisfies
$|\omega(X)|\le N_t$.
Hence $\tau_s^{\mathrm{typ}}(G_0)\le N_t$.
\end{lemma}

\begin{proof}
The \emph{continuing spine} follows the unique
non-wrapper child of each binary rule.
The wrapper child generates exactly one terminal
and never continues the spine. Unit rules add
no terminals; lexical and nullable rules end
the spine. Choose a derivation of $\omega(X)$
with the fewest nonterminal occurrences.
A repeated \emph{typed} spine symbol permits
grafting its lower subderivation at the upper
occurrence, preserving a valid typed derivation.
A segment containing a binary rule would strictly
shorten $\omega(X)$; a unit-only segment would
decrease the number of nodes. The former contradicts the minimal length of $\omega(X)$;
the latter contradicts the choice of a derivation with the fewest
nonterminal occurrences.
At most $N_t$ spine symbols occur, and each contributes
at most one terminal, including the final lexical rule.
Therefore $|\omega(X)|\le N_t$. Wrapper symbols
also satisfy this estimate.
\end{proof}

\begin{theorem}[polynomial characteristic data for linear targets]
\label{thm:linear-poly}
For fixed $s$, every $L\in\Clin{s}$ represented by a linear CFG $G$ has
a characteristic sample for $\mathsf{Rec}_s$ whose $\|\cdot\|$-size is
polynomial in $|G|$.
\end{theorem}

\begin{proof}
For $L=\emptyset$ and $L=\{\lambda\}$ use the samples from
Section~\ref{sec:typing}.  Otherwise take $G_0$ from
Proposition~\ref{prop:linear-normal}.  Lemma~\ref{lem:linear-short} gives
$\tau_s^{\mathrm{typ}}(G_0)\le N_t\le
|V_0|\,|\mathsf S^\sharp|$, and
Corollary~\ref{cor:typed-thickness-data} yields a characteristic sample
polynomial in $|G_0|\le q_{\mathrm{lin}}(|G|)$. More explicitly, the bound in
Lemma~\ref{lem:typed-thickness-bound} gives
$\|\WitnessSet(\widetilde G_0)\|=O(|G_0|^3|\mathsf S^\sharp|^4)$;
the polynomial dependence on the original $|G|$ also uses
the normalization bound $q_{\mathrm{lin}}$.
\end{proof}

Together with Theorem~\ref{thm:poly-build}, this gives
polynomial-time batch reconstruction and polynomial characteristic data
relative to a chosen linear CFG for targets in $\Clin{s}$.
This does not contradict de~la~Higuera's negative result for
the unrestricted class of linear languages
\cite[Theorem~2]{higuera1997}: our theorem concerns only the
fixed-$s$ substitutable subclass.  Moreover, no incremental
polynomial-data guarantee for $\mathsf{Learn}_s$ is asserted.

\paragraph{A nonregular linear \texorpdfstring{fixed-$s$}{fixed-s} example.}

The preceding theorem is not confined to regular targets, nor to the
Clark--Eyraud or fixed-window classes.  Define
\[
\begin{split}
L_{\pm,e}:={}&
\{a^{2m}cb^{2m}:m\ge0\}
\cup
\{a^{2m+1}db^{2m+1}:m\ge0\}\\
&\cup
\{a^neb^n:n\ge0\}.
\end{split}
\]
It is generated by the linear grammar
\[
\begin{array}{lll}
S\to U_{\mathrm{ev}}\mid U_e, &
U_{\mathrm{ev}}\to c\mid aU_{\mathrm{odd}}b, &
U_{\mathrm{odd}}\to d\mid aU_{\mathrm{ev}}b,\\[1mm]
&&U_e\to e\mid aU_eb.
\end{array}
\]

The subscripts $\mathrm{ev}$ and $\mathrm{odd}$ track the parity
of enclosing $a,b$-pairs from the start, not the length parity
of the corresponding nonterminal's own yield.

Call $c,d,e$ the \emph{center symbols}.

\begin{proposition}\label{prop:linear-separator-example}
The language $L_{\pm,e}$ is linear, nonregular, and belongs to
$\Clin{s}$ for some homomorphism $s$ into a finite semigroup.
It is neither Clark--Eyraud substitutable nor
$(k,\ell)$-substitutable for any fixed $k,\ell\ge0$.
\end{proposition}

\begin{proof}
Linearity follows from the displayed grammar. To establish finite-typed
substitutability, put
\[
 L_{\mathrm{all}}:=\{a^n\xi b^n:n\ge0,\ \xi\in\{c,d,e\}\},\qquad
 R_{\mathrm{reg}}:=(aa)^*c(bb)^*\cup a(aa)^*d\,b(bb)^*\cup a^*eb^*.
\]
The linear language $L_{\mathrm{all}}$ is Clark--Eyraud substitutable
under the nonempty-factor convention: a shared context uniquely determines
a center-free factor, while the distribution of a factor
$a^i\xi b^j$ containing one center depends only on $i-j$.
Thus $L_{\mathrm{all}}\in\Clin{s_0}$ for the trivial typing $s_0$.
Since $R_{\mathrm{reg}}$ is regular and $L_{\pm,e}=L_{\mathrm{all}}\cap R_{\mathrm{reg}}$,
let $g:\Sigma^*\to M$ be a finite-monoid morphism recognizing
$R_{\mathrm{reg}}$. Put $\mathsf T=g(\Sigma^+)$ and
$t=g|_{\Sigma^+}:\Sigma^+\twoheadrightarrow\mathsf T$.
Because $\lambda\notin R_{\mathrm{reg}}$, a suitable
$\mathsf{Acc}_t\subseteq\mathsf T\subseteq\mathsf T^\sharp$ satisfies
$R_{\mathrm{reg}}=\widehat t^{-1}(\mathsf{Acc}_t)$.
Proposition~\ref{prop:finite-info-closure}(ii), together with closure
of linear languages under regular intersection, then gives
$L_{\pm,e}\in\Clin{s_0\times t}$.

Moreover, $L_{\pm,e}\cap a^*eb^*=\{a^neb^n:n\ge0\}$ proves nonregularity.
The factors $c,e$ share the empty context, but
$(a,b)\in\D_{L_{\pm,e}}(e)\setminus\D_{L_{\pm,e}}(c)$;
hence $L_{\pm,e}$ is not Clark--Eyraud substitutable.
Finally, fix $k,\ell\ge0$ and put
$m_0:=\max\{1,k+\ell\}$,
$w_e:=a^keb^\ell$, and $w_c:=a^kcb^\ell$.
The factors $w_e,w_c$ have the same length-$k$ prefix and
length-$\ell$ suffix and share the context
$(a^{2m_0-k},b^{2m_0-\ell})$, whereas
$(a^{2m_0+1-k},b^{2m_0+1-\ell})
\in\D_{L_{\pm,e}}(w_e)\setminus\D_{L_{\pm,e}}(w_c)$.
The \hyperref[crit:kl]{fixed-window counterexample criterion} gives
$L_{\pm,e}\notin\mathsf{KL}$.
\end{proof}

\paragraph{A concrete four-element typing.}
The regular-filter argument establishes the required existence of a
finite typing. In fact, a four-element typing already suffices.
Let $M_{\pm,e}:=\{1,\kappa_{cd},\kappa_e,0\}$ be the monoid with identity
$1$ and $\mu\nu=0$ for all
$\mu,\nu\in\{\kappa_{cd},\kappa_e,0\}$, and define
$s_{\pm,e}:\{a,b,c,d,e\}^+\to M_{\pm,e}$ by
\begin{equation}\label{eq:lpm-s}
 s_{\pm,e}(a)=s_{\pm,e}(b)=1,\qquad
 s_{\pm,e}(c)=s_{\pm,e}(d)=\kappa_{cd},\qquad
 s_{\pm,e}(e)=\kappa_e.
\end{equation}
Then $L_{\pm,e}$ is $\sim_{s_{\pm,e}}$-substitutable.
Indeed, two center-free nonempty factors sharing a context
must coincide. The distribution of $a^ieb^j$ depends only on $i-j$,
while that of $a^i\xi b^j$ for $\xi\in\{c,d\}$ depends only on
$(i-j,\operatorname{par}(\xi)-i\bmod2)$, where
$\operatorname{par}(c)=0$ and $\operatorname{par}(d)=1$.
A shared context forces these invariants to agree, and equal
$s_{\pm,e}$-types distinguish $e$-factors from $c/d$-factors;
thus equal-type factors with a common context have the same distribution.

Thus Theorem~\ref{thm:linear-poly} applies to a nonregular linear language
outside $\mathsf{KL}$. The displayed grammar is even-linear, so no
separation from Takada's control-set framework is claimed.

\paragraph{Finite-state control induced by yield typing.}
A single finite-state construction covers both the
original typing $s$ and its completion $\widehat s$.
Let $t:\Sigma^*\to T$ be any finite-monoid
morphism, and label each production of a linear
grammar $G=(V,\Sigma,P,S)$ distinctly over
$\mathsf{ProdLab}$.  Write $A\Rightarrow_G^\alpha w$
for a derivation with production-label word $\alpha$.
For $\mathsf{Acc}_{\mathrm{ctrl}}\subseteq T$, put
\[
\mathsf{Ctrl}_{t,\mathsf{Acc}_{\mathrm{ctrl}}}(G):=
\{\alpha\in\mathsf{ProdLab}^*:S\Rightarrow_G^\alpha w
\text{ for some }w\in\Sigma^*,\
t(w)\in\mathsf{Acc}_{\mathrm{ctrl}}\}.
\]
This is regular: an automaton with states
$V\times T\times T$ and an accepting state starts
at $(S,1_T,1_T)$, updates
$(A,\mu_L,\mu_R)\mapsto
(B,\mu_Lt(u),t(v)\mu_R)$ for each
$A\to uBv$, and accepts a terminal rule
$A\to w$ iff $\mu_Lt(w)\mu_R\in
\mathsf{Acc}_{\mathrm{ctrl}}$.
The invariant is $(A,t(u),t(v))$ for a
sentential form $uAv$; thus the controlled
grammar generates $L(G)\cap
t^{-1}(\mathsf{Acc}_{\mathrm{ctrl}})$.

The semigroup typing $s$ is used through its monoid extension:
taking $t=\widehat s$, start symbol $A$, and
$\mathsf{Acc}_{\mathrm{ctrl}}=\{\mu\}$
selects precisely
$L_G(A)\cap \widehat s^{-1}(\{\mu\})$,
the fiber in Proposition~\ref{prop:typed-core},
including the nullable type $\mathbf{1}_{\sharp}$.
This is a finite-state bridge to Takada's
regular control languages~\cite{takada1995};
we do not claim arbitrary Takada control sets
arise from yield typings.

\section{Expressiveness and Boundaries}
\label{sec:boundaries}

We now locate the framework between the fixed-window hierarchy and broader
context-free families.

\subsection{A nonlinear target inside \texorpdfstring{$\mathsf{RS}$}{RS}}

Put $\Delta:=\{a^nb^n:n\ge0\}$ and, for
$z\in\{a,b\}^*$, let $\beta(z):=|z|_a-|z|_b$.
The \emph{height} after reading a prefix $p$ is $\beta(p)$.
Membership in $\Delta^*$ is characterized by the total
height and the heights at the occurrences of $ba$:
\begin{equation}\tag{$\star$}
w\in\Delta^*
\iff \beta(w)=0\quad\text{and}\quad
\beta(pb)=0\ \text{whenever }w=pba p'.
\end{equation}
Indeed, a word in $\Delta^*$ is a product of blocks
$a^nb^n$, whose nontrivial boundaries are precisely its
occurrences of $ba$ and have height zero.
Conversely, cut at every occurrence of $ba$. The resulting
factors contain no $ba$ and therefore belong to $a^*b^*$;
the right-hand conditions force each factor to start and
end at height zero. Thus every factor is $a^nb^n$.

For a word $x$, define its \emph{boundary-compatible
entry-offset set} by
\[
\mathcal Q(x):=
\{\delta\in\mathbb Z:
 \delta+\beta(pb)=0\text{ for every factorization }x=pba p'\}.
\]
These are formal integer offsets, not necessarily reachable
nonnegative stack heights. If $x$ contains no $ba$, then
$\mathcal Q(x)=\mathbb Z$; otherwise
$|\mathcal Q(x)|\le1$. For example,
$\mathcal Q(ab)=\mathcal Q(bb)=\mathbb Z$,
$\mathcal Q(ba)=\{1\}$,
$\mathcal Q(aaba)=\{-1\}$, and
$\mathcal Q(baaba)=\varnothing$.

For nonempty $x$, every occurrence of $ba$ in $uxv$
lies inside $u$, $x$, or $v$, or straddles one of the
two concatenation boundaries. Hence \textup{($\star$)}
gives the following exact membership criterion:
\begin{equation}\tag{$\star\star$}
\begin{gathered}
uxv\in\Delta^*
\iff
\beta(uxv)=0,\quad
0\in\mathcal Q(u),\quad
\beta(u)\in\mathcal Q(x),\quad
\beta(ux)\in\mathcal Q(v),\\
\bigl(\operatorname{lst}(u),\operatorname{fst}(x)\bigr)=(b,a)
\ \Longrightarrow\ \beta(u)=0,\\
\bigl(\operatorname{lst}(x),\operatorname{fst}(v)\bigr)=(b,a)
\ \Longrightarrow\ \beta(ux)=0.
\end{gathered}
\end{equation}
Here $\operatorname{fst}$ and $\operatorname{lst}$ denote the
first and last letters of a nonempty word, with the value
$\bot$ for the empty word. Thus membership of $uxv$
depends on $x$ only through $\beta(x)$,
$\mathcal Q(x)$, and its first and last letters.

For nonempty $w$, let $s_\star(w)$ record its first and
last letters and whether it contains an occurrence of $ba$.
The summary of $xy$ is determined by those of $x$ and $y$:
the outer letters are inherited, and a new $ba$ can appear
only across a boundary $b|a$. Thus $\ker s_\star$ is a
semigroup congruence, and $s_\star$ is a surjective
homomorphism onto its finite positive-word image,
containing seven elements: the combination of first letter $b$,
last letter $a$, and no occurrence of $ba$ is impossible.

\begin{proposition}[a nonlinear fixed-$s_\star$ target]
\label{prop:nonlinear-rs-example}
The language $\Delta^*$ is deterministic context-free, nonregular, and
nonlinear, and it is $\sim_{s_\star}$-substitutable. Moreover it is not
$(k,\ell)$-substitutable for any fixed $k,\ell\ge0$. Hence
$\Delta^*\in\mathsf{RS}_{s_\star}(\{a,b\})\cap\mathrm{CFL}$ and
$\Delta^*\in\mathsf{RS}(\{a,b\})\setminus\mathsf{KL}(\{a,b\})$.
\end{proposition}

\begin{proof}
A deterministic pushdown automaton recognizes $\Delta^*$ by
tracking height, rejecting a $b$ read at height zero and
an $a$ immediately after a $b$ when the height is nonzero,
and accepting at height zero; this follows from
\textup{($\star$)}. The grammar
$S\to TS\mid\lambda$, $T\to aTb\mid\lambda$
also generates $\Delta^*$.
Nonregularity follows from
$\Delta^*\cap a^*b^*=\Delta$. For nonlinearity,
linear languages are closed under regular intersection,
whereas $\Delta^*\cap a^*b^*a^*b^*=\Delta\Delta$
is not linear
\cite[Section~6.1, before Theorem~6.1]{autebert1997}.

For substitutability, let nonempty $x,y$ have equal
$s_\star$-types and a common context $(u,v)$.
From $uxv,uyv\in\Delta^*$ and \textup{($\star\star$)},
we obtain $\beta(x)=\beta(y)$ and
$\beta(u)\in\mathcal Q(x)\cap\mathcal Q(y)$.
If neither $x$ nor $y$ contains $ba$, then
$\mathcal Q(x)=\mathcal Q(y)=\mathbb Z$.
Otherwise both contain $ba$, and each offset set
has at most one element; hence
$\mathcal Q(x)=\{\beta(u)\}=\mathcal Q(y)$.
Since $x,y$ also have equal first and last letters,
\textup{($\star\star$)} implies, for all $u',v'$,
\[
u'xv'\in\Delta^*\iff u'yv'\in\Delta^*.
\]
Therefore $\D_{\Delta^*}(x)=\D_{\Delta^*}(y)$.

For the fixed-window separation, fix $k,\ell\ge0$
and set $m_0:=\max\{1,k+\ell\}$,
$x:=a^{m_0}b^{m_0}$, and
$y:=a^{m_0}b^{m_0}a^{m_0}b^{m_0}$.
The words share the same length-$k$ prefix and
length-$\ell$ suffix, and the empty context.
But $axb\in\Delta^*$, while the central $ba$ in
$ayb$ has height one immediately after its $b$;
thus $ayb\notin\Delta^*$ by \textup{($\star$)}.
The fixed-window counterexample criterion excludes
every $(k,\ell)$.
\end{proof}

\subsection{A capped regular counter family}

Yoshinaka's regular language
$L_{\mathrm{Yreg}}=ae^*ce^*a\cup ae^*de^*a\cup be^*ce^*b$
already gives a fixed-window separation
\cite[Proposition~1]{yoshinaka2008}; capped counters contrast
with the uncapped obstruction below.

Let $\Sigma_c:=\{\uparrow,\downarrow,\#\}$.  For
$w=c_1\cdots c_n\in\Sigma_c^*$, let $\mathrm{ht}_i(w)$
be the height after $i$ letters, with $\mathrm{ht}_0(w)=0$, changing
by $+1$ on $\uparrow$, $-1$ on $\downarrow$, and resetting to $0$ on $\#$.
For $B\ge1$, put
\[
\mathrm{CTR}_B:=
\{w\in\Sigma_c^*:0\le\mathrm{ht}_i(w)\le B
\text{ for all }0\le i\le|w|\}.
\]

For every $B\ge1$, $\mathrm{CTR}_B$ is regular: a deterministic
finite automaton records
heights $0,\ldots,B$, uses one sink outside the range, and resets to $0$
on $\#$.  Hence $\mathrm{CTR}_B\in\mathsf{RS}(\Sigma_c)$ by
Proposition~\ref{prop:regular-auto}.

\begin{theorem}\label{thm:ctr-non-kl}
For every $B\ge2$, $\mathrm{CTR}_B$ is not
$(k,\ell)$-substitutable for any $k,\ell\ge0$.
\end{theorem}
\begin{proof}
Fix $k,\ell$ and put
$x:=(\uparrow\downarrow)^k\uparrow^{B-1}\#^\ell$ and
$y:=(\uparrow\downarrow)^k\uparrow^B\#^\ell$.  They are long
enough for the fixed-window criterion, have the same length-$k$ prefix and
length-$\ell$ suffix, and both lie in $\mathrm{CTR}_B$,
so they share the empty context. Yet
$\uparrow x\in\mathrm{CTR}_B$ while
$\uparrow y\notin\mathrm{CTR}_B$, the latter reaching height
$B+1$.  Hence their distributions differ, and the criterion applies.
\end{proof}

The bound $B\ge2$ is sharp\footnote{Indeed,
$\mathrm{CTR}_1$ is $(1,1)$-substitutable. Membership is characterized
by the initial symbol not being $\downarrow$ and the absence
of the length-$2$ factors $\uparrow\uparrow$,
$\downarrow\downarrow$, and $\#\downarrow$. Thus, for nonempty factors occurring in words of $\mathrm{CTR}_1$
with a common context, equality of their first and last symbols
ensures equal membership under every context.}.
Let $h_B:\Sigma_c^*\to M_B$ be the
DFA transition-monoid morphism recognizing
$\mathrm{CTR}_B$. Its positive restriction
$s_B:=h_B|_{\Sigma_c^+}$ is a finite-semigroup
typing witnessing $\mathrm{CTR}_B\in\Ccf{s_B}$;
only $B\ge2$ lies outside the fixed-window hierarchy.

\subsection{Natural deterministic context-free languages outside \texorpdfstring{$\mathsf{RS}$}{RS}}\label{sec:outside-rs}

\begin{lemma}[finite-semigroup obstruction]\label{lem:finite-monoid-obstruction}
Let $L\subseteq\Sigma^*$.  Suppose an infinite
$\Xi\subseteq\Sigma^+$ satisfies, for all distinct $x,y\in\Xi$,
$\D_L(x)\cap\D_L(y)\ne\emptyset$ and
$\D_L(x)\ne\D_L(y)$.  Then $L\notin\mathsf{RS}(\Sigma)$.
\end{lemma}
\begin{proof}
Any finite $s$ identifies two elements of $\Xi$; their common context and
unequal distributions contradict $\sim_s$-substitutability.
\end{proof}

Removing the upper cap gives the deterministic context-free language
\[
\mathrm{CTR}:=
\{w\in\Sigma_c^*: \mathrm{ht}_i(w)\ge0
\text{ for all }0\le i\le|w|\},
\]
accepted by the evident deterministic pushdown counter with reset.

\begin{corollary}[uncapped counter obstruction]
$\mathrm{CTR}\notin\mathsf{RS}(\Sigma_c)$.
\end{corollary}
\begin{proof}
Use $\Xi=\{\uparrow^i:i\ge1\}$.  For $i<j$,
$(\lambda,\downarrow^i)$ is common, while
$(\lambda,\downarrow^j)$ is a context only of $\uparrow^j$.
\end{proof}

Define the one-bracket Dyck language by
\[
 D_1:=\{w\in\{a,b\}^*: |w'|_a\ge|w'|_b
 \text{ for every prefix }w',\ |w|_a=|w|_b\}.
\]
It is a standard deterministic context-free language
\cite{ginsburg-greibach1966}.
Here $D_1$ denotes the whole Dyck language, written $D_1^*$
in the convention of Autebert--Berstel--Boasson~\cite{autebert1997}.

\begin{corollary}\label{cor:dyck-not-rs}
$D_1\notin\mathsf{RS}(\{a,b\})$.
\end{corollary}
\begin{proof}
Use $\Xi=\{b^ia^i:i\ge1\}$.  For $i<j$,
$(a^j,b^j)$ is common, whereas $(a^i,b^i)$ is a context only of
$b^ia^i$; for $b^ja^j$ the prefix $a^ib^j$ underflows.
\end{proof}

Thus unbounded stack behavior alone is not the obstruction: the nonregular
one-turn language $L_{\pm,e}$ lies in $\mathsf{RS}$, but an infinite family
of mutually comparable factors with distinct contextual behavior cannot be
separated by any finite $s$.

\begin{lemma}[fixed-word quotient]\label{lem:rs-fixed-quotient}
If $L$ is $\sim_s$-substitutable and $z\in\Sigma^*$, then
$L/z:=\{w:wz\in L\}$ is $\sim_s$-substitutable.
\end{lemma}
\begin{proof}
A common context $(u,v)$ in $L/z$ becomes $(u,vz)$ in $L$;
equality of distributions in $L$ then transfers every
context $(u',v')$ by appending $z$.
\end{proof}

The language generated by $S\to aSS\mid b$ is the
\emph{{\fontencoding{T1}\selectfont\L{}}ukasiewicz language}.
Yoshinaka~\cite[Section~3]{yoshinaka2008} already observed that this language
is not $(k,\ell)$-substitutable for any fixed $k,\ell$.  In the notation used
here, $L_{\mathrm{Luk}}=D_1b$~\cite{autebert1997}, and it is deterministic
context-free~\cite{ginsburg-greibach1966}.  The quotient argument gives the
stronger conclusion that it lies outside the present finite-typing union: if
$L_{\mathrm{Luk}}\in\mathsf{RS}(\{a,b\})$, then
Lemma~\ref{lem:rs-fixed-quotient} would put
$D_1=L_{\mathrm{Luk}}/b$ in $\mathsf{RS}$, contradicting
Corollary~\ref{cor:dyck-not-rs}.

\subsection{Comparison with Clark's congruential family}

Clark~\cite{clark2010congruential} allows a finite initial set
$I\subseteq V$; write $L_G[I]:=\bigcup_{A\in I}L_G(A)$.  Such a grammar is
\emph{congruential} if each $L_G(A)$ lies in one syntactic-congruence class
of $L_G[I]$.  Let $\mathsf{CONG}(\Gamma)$ be the languages over $\Gamma$
generated by such grammars.  Clark already places substitutable and
$(k,\ell)$-substitutable languages in this family; the following extends
the containment to every fixed-$s$ slice.

\begin{proposition}[comparison with Clark's congruential family]
\label{prop:clark-congruential-comparison}
For every finite alphabet $\Gamma$ and finite-semigroup typing
$s:\Gamma^+\to\mathsf S$,
$\Ccf{s}\subseteq\mathsf{CONG}(\Gamma)$.  Consequently
$\mathsf{RS}(\Gamma)\cap\mathrm{CFL}
\subseteq\mathsf{CONG}(\Gamma)$, and for $\Gamma=\{a,b\}$ the
inclusion is proper.
\end{proposition}

\begin{proof}
For nonempty $L\in\Ccf{s}$, take any CFG for $L$
and its trimmed $\mathsf S^\sharp$-typed refinement $\widetilde G$.
The yields of each retained typed symbol $A_\mu$
have equal $\widehat s$-type and share a terminal reaching
context.  Lemma~\ref{lem:fresh-identity} implies
that they lie in one syntactic-congruence class.
This argument includes nullable symbols.

Delete the auxiliary $\widetilde S$ and use its
typed successors $(S_0)_\mu$ as a finite initial set.
The type $\mathbf{1}_{\sharp}$ already accounts for $\lambda$
when it belongs to $L$; no extra initial symbol
is necessary.  A productionless initial symbol
handles the empty language.  Taking the union
over all finite typings gives the inclusion.

For properness over $\{a,b\}$, use the Dyck example also discussed by
Clark~\cite{clark2010congruential}.  Every $x\in D_1$ satisfies
$uxv\in D_1$ iff $uv\in D_1$, so $x\equiv_{D_1}\lambda$.  Hence
$S\to aSbS\mid\lambda$ is congruential, while
Corollary~\ref{cor:dyck-not-rs} gives
$D_1\notin\mathsf{RS}(\{a,b\})$.
\end{proof}

\section{Conclusion}

Qualitative learnability is insensitive to grammatical normal
forms, whereas quantitative data bounds reveal the algebraic
structure of the typing. The exponential separation concerns
typed thickness across BUNF presentations and characteristic
samples for set-driven operators, not convergence of arbitrary
Gold learners. The window--group hypothesis is sufficient,
not claimed necessary; characterizing
$\mathsf{RS}\cap\mathrm{CFL}$ remains open.

\section*{Data availability}
No empirical data were used for the research described in this article.

\section*{Declaration of competing interest}
The author declares no known competing financial interests or personal
relationships that could have influenced the work reported in this paper.

\section*{Funding}
This research received no specific grant from any funding agency in the
public, commercial, or not-for-profit sectors.

\section*{Acknowledgments}
The author thanks Makoto Kanazawa and Ryo Yoshinaka for guidance,
encouragement, and help with relevant literature, and the two anonymous
reviewers for careful and constructive reports that substantially improved
the paper, in particular by identifying problems in the submitted learning
algorithm and by prompting the direct two-word witness in
Theorem~\ref{thm:ctr-non-kl}.
The author also thanks Alexander Clark for valuable advice at
ICGI 2014 during the early development of this work.
Any remaining errors are the author's responsibility.

\appendix
\section{Fixed-window witness bounds and thickness-preserving normalization}
\label{app:fixed-window-bounds}

\subsection{Proof of Lemma~\ref{lem:window-typed-yield}}
\begin{proof}[Proof of Lemma~\ref{lem:window-typed-yield}]
Put $m:=k+\ell$, $q:=|\mathsf H|$, $N:=|V|$, and
$\bar\tau:=\tau_G+1$.
Fix $A\in V$, possibly $S_0$, and a retained
typed copy $A_\mu$ with $\mu\ne\mathbf{1}_{\sharp}$;
the copy $A_{\mathbf{1}_{\sharp}}$ yields only $\lambda$.
Let $\eta$ be the $(s_{k,\ell}\times g)$-value of some
nonempty $w_0\in L_G(A)$ of $\widehat s$-type $\mu$.
Among all derivation trees from $A$ yielding a nonempty word
of $(s_{k,\ell}\times g)$-value $\eta$, choose $T$, with
yield $w$, minimizing first $|w|$ and then the number
of nonterminal nodes. The kernel hypothesis gives
$\widehat s(w)=\mu$, so
$|\omega(A_\mu)|\le|w|$ by
Proposition~\ref{prop:typed-core}.
The bound is immediate if $|w|<m$; hence assume
$|w|\ge m$.

The first $k$ and last $\ell$ positions of $w$ are
disjoint. Protect them and choose one arbitrary position
of $w$ as a \emph{nonemptiness anchor}, which may already
belong to a protected window. Let $\mathrm{Prot}$ be the
union of these positions. Then
$1\le d:=|\mathrm{Prot}|\le m+1$.
In particular, when $m=0$ the anchor still prevents
the word from becoming empty. We use the following
minimality consequence:
\begin{itemize}[leftmargin=2.2em]
\item[$(\ast)$]
If $w'$ arises by replacing pairwise disjoint factors of
$w$ whose occurrences contain no protected position,
and $g(w')=g(w)$, then $w'$ is nonempty and has the same
$(s_{k,\ell}\times g)$-value $\eta$ as $w$.
Thus a derivation tree from $A$ for such a $w'$ cannot
have shorter yield, or the same yield length and
fewer nonterminal nodes, than $T$.
\end{itemize}
Indeed, the first $k$ and last $\ell$ protected letters
remain the corresponding windows, and
$|w'|\ge d\ge1$.

Consider only \emph{nonterminal occurrences} in $T$,
including those arising from unit and nullable productions.
For each such node $z$, let $X_z\in V$ be its label,
$y_z$ its possibly empty terminal yield, and
$P_z\subseteq\mathrm{Prot}$ the protected positions
in its yield span. Call $z$ \emph{essential} if
$P_z\ne\emptyset$ or $|y_z|>\bar\tau$.

\emph{Depth.}
If a proper descendant $z'$ of $z$ has the same
triple $(P_z,X_z,g(y_z))$ as $z$, graft the subtree
at $z'$ in place of the subtree at $z$.
The removed factors are disjoint from $\mathrm{Prot}$,
and the group value is preserved, so $(\ast)$ applies.
The grafted tree has no longer yield and strictly
fewer nonterminal nodes, a contradiction.
Along any path the sets $P_z$ are nested and
take at most $d+1$ values. Hence no path contains
more than $(d+1)Nq\le(m+2)Nq$ nonterminal nodes.

\emph{Width.}
If a level contained $d+q$ essential nodes,
their disjoint yield spans would leave at least
$q$ nodes avoiding $\mathrm{Prot}$. Choose $q$
such nodes, in left-to-right order, with yields
$a_1,\ldots,a_q$ of length greater than $\bar\tau$.
For each, choose a word $b_i$ generated by its
nonterminal label with $|b_i|\le\tau_G$;
$b_i=\lambda$ is permitted.
Write $w=c_0a_1c_1\cdots a_qc_q$.
For $1\le j\le q$, put
$r_j:=g(a_1c_1\cdots a_jc_j)$ and
$t_j:=g(b_1c_1\cdots b_jc_j)$, and define
$x_j:=t_jr_j^{-1}$, with $r_0=t_0=x_0:=1_{\mathsf H}$.
Two of $x_0,\ldots,x_q$ coincide, say
$x_i=x_j$ with $i<j$. Write the ordered interval
products as $r_j=r_i r_{i:j}$ and
$t_j=t_i t_{i:j}$. From
$t_i r_i^{-1}=t_i t_{i:j}r_{i:j}^{-1}r_i^{-1}$,
cancellation in this order gives
$t_{i:j}=r_{i:j}$. Thus the corresponding
interval words have the same $g$-value.
Replace the corresponding pairwise disjoint
subtrees by derivations of $b_{i+1},\ldots,b_j$.
This preserves the group value and all protected
positions, but strictly shortens the yield,
contradicting $(\ast)$. No commutativity
of $\mathsf H$ is used. Therefore every
level has at most $d+q-1\le m+q$
essential nodes.

\emph{Counting.}
The essential nonterminal nodes are closed under
ancestors and include the root, since
$d\ge1$. Their total number is at most
$E:=(m+q)(m+2)Nq$.
As $G$ is in BUNF, at most $2E$ maximal
nonessential subtrees hang below them, each
yielding at most $\bar\tau$ letters.
Every remaining terminal letter is emitted
by an essential lexical node, hence is
protected, because a lexical node has yield
length $1\le\bar\tau$.
There are at most $d$ such letters. Thus
\[
|w|\le d+2E\bar\tau
\le m+1+
[1+2(m+q)(m+2)Nq](\tau_G+1).
\qedhere
\]
\end{proof}

\subsection{Thickness-preserving BUNF normalization}
\label{app:thick-bunf}
\begin{proof}[Proof of Proposition~\ref{prop:thick-bunf-normal}]
Let $G_*$ be reduced and put $n:=|G_*|$,
$\bar\tau:=\tau_{G_*}+1$.
First isolate terminals in right-hand sides
of length at least two, using wrappers
$W_a\to a$, and then binarize.
Retain all nullable and unit rules.
These standard operations preserve all
original nonterminal languages
\cite[Section~3.1]{autebert1997}
\cite[Sections~7.1 and~7.4.2]{hopcroft-motwani-ullman2007}.
Each auxiliary binarization symbol is fresh
for its original production and represents a
contiguous block of that production's right-hand
side, containing at most $O(n)$ original symbols.
Substituting shortest yields gives a terminal
word of length $O(n\bar\tau)$.
The original nonterminal languages and their
shortest yields are unchanged.
Trim unreachable or unproductive symbols.
The resulting BUNF grammar is equivalent,
has polynomial size and
$\tau_G=O(n(\tau_{G_*}+1))$.
\end{proof}

\section{Proof of the linear-spine BUNF normalization}
\label{app:linear-normalization}
\begin{proof}[Proof of Proposition~\ref{prop:linear-normal}]
A linear CFG has rules $A\to uBv$
or $A\to w$, for terminal words
$u,v,w\in\Sigma^*$.
Retain $A\to B$ and $A\to\lambda$.
No nullable or unit elimination is needed.

For each $a\in\Sigma$ introduce a fresh
terminal wrapper $W_a\to a$, and choose
all other auxiliary spine symbols freshly
per production. For a rule
$A\to a_1\cdots a_m B b_1\cdots b_n$
with $m+n>0$, form the chains
\[
\begin{aligned}
C^{\mathrm R}_0&:=B,&
C^{\mathrm R}_j&\to C^{\mathrm R}_{j-1}W_{b_j}
  &&(1\le j\le n),\\
C^{\mathrm L}_0&:=C^{\mathrm R}_n,&
C^{\mathrm L}_i&\to W_{a_{m-i+1}}C^{\mathrm L}_{i-1}
  &&(1\le i\le m).
\end{aligned}
\]
Identify $C^{\mathrm R}_n=A$ when $m=0$,
and $C^{\mathrm L}_m=A$ otherwise.
For terminal-only $A\to a_1\cdots a_k$,
keep $k=0,1$ unchanged. If $k\ge2$,
let $C^{\mathrm T}_1=A$ and use fresh
$C^{\mathrm T}_2,\ldots,C^{\mathrm T}_k$ with
\[
C^{\mathrm T}_i\to W_{a_i}C^{\mathrm T}_{i+1}
\ (1\le i<k),\qquad C^{\mathrm T}_k\to a_k.
\]
Every new binary rule has one wrapper
child producing exactly one terminal,
and one continuing spine child.
The original languages are preserved.
Trim to obtain a reduced BUNF grammar
of polynomial encoded size.
\end{proof}

\section{Substitutability of the bracketed doubling family}
\label{app:bracketed-substitutability}

Put $L_0=\{\mathtt a\}$ and
$L_i=\lceil{}L_{i-1}\rceil\cup\lfloor{}L_{i-1}L_{i-1}\rfloor$ for $i\ge1$;
thus $L_{G_n}(E_i)=L_i$ for $0\le i\le n$, and $L_n=L(G_n)$.
Fix $n\ge1$. We prove
Proposition~\ref{prop:typed-thickness-gap}(i) with a deterministic
top-down parser for $G_n$ whose transitions are injective for
each fixed input letter.

\paragraph{The parser.}
A \emph{stack} is a word over $E_0,\ldots,E_n,\rceil,\rfloor$, with its
top on the left; $\lambda$ is the empty stack. For every stack $\psi$
and $1\le i\le n$ the moves are
\[
\begin{aligned}
E_0\psi&\xrightarrow{\mathtt a}\psi,&
E_i\psi&\xrightarrow{\lceil}E_{i-1}\rceil\psi,&
E_i\psi&\xrightarrow{\lfloor}E_{i-1}E_{i-1}\rfloor\psi,\\
\rceil\psi&\xrightarrow{\rceil}\psi,&
\rfloor\psi&\xrightarrow{\rfloor}\psi,
\end{aligned}
\]
and there are no others; in particular, no move leaves $\lambda$.
Write $\sigma\xrightarrow{w}\theta$ if reading $w$ from $\sigma$ ends
in $\theta$. The parser starts with the stack $E_n$ and accepts $w$ if
$E_n\xrightarrow{w}\lambda$; Lemma~\ref{lem:app-recognition} shows that
it accepts exactly $L_n$. We use three properties.

\begin{enumerate}[label=\textup{(P\arabic*)},leftmargin=2.6em]
\item \emph{Determinism and letterwise injectivity.}
For each input letter $c$, the transition relation
defines a partial injective map on stacks:
there is at most one $c$-successor of each stack,
and two $c$-transitions with the same target
have the same source.
For opening brackets, the output reveals the
expanded $E_i$ and the untouched suffix;
for $\mathtt a$, $\rceil$, and $\rfloor$,
the input letter identifies the popped top symbol.
Consequently every word acts as a partial injection
on stacks: if $\sigma\xrightarrow{w}\theta$ and
$\sigma\xrightarrow{w}\theta'$, then $\theta=\theta'$;
if $\sigma\xrightarrow{w}\theta$ and
$\sigma'\xrightarrow{w}\theta$, then $\sigma=\sigma'$.

\item \emph{Weight.} Give $E_i$ weight $4^i$ and each closing marker
weight $1$, and let $\operatorname{wt}(\sigma)$ be the total weight of
a stack $\sigma$. Every move lowers $\operatorname{wt}$:
each pop removes weight $1$, and
$4^i>2\cdot4^{i-1}+1$ for $i\ge1$.
Hence $\sigma\xrightarrow{w}\sigma$ only for $w=\lambda$.

\item \emph{Locality.} Moves inspect and rewrite only the top symbol,
and closing markers are only popped. Hence
$\sigma\xrightarrow{w}\theta$ implies
$\sigma\psi\xrightarrow{w}\theta\psi$. Conversely, let
$\sigma_1\ne\lambda$ and $\sigma_1\psi\xrightarrow{w}\theta$. Then
either
\begin{enumerate}[label=\textup{(\alph*)},nosep]
\item the run visits $\psi$: splitting at the first visit gives
$w=w_1w_2$ with $\sigma_1\xrightarrow{w_1}\lambda$ and
$\psi\xrightarrow{w_2}\theta$, so
$\operatorname{wt}(\theta)\le\operatorname{wt}(\psi)$; or
\item it does not: $\theta=\theta_1\psi$ with
$\sigma_1\xrightarrow{w}\theta_1\ne\lambda$, so
$\operatorname{wt}(\theta)>\operatorname{wt}(\psi)$, and $\theta_1$ ends
with the last symbol of $\sigma_1$ whenever that symbol is a closing
marker.
\end{enumerate}
Thus, for the fixed suffix $\psi$, the comparison of
$\operatorname{wt}(\theta)$ with $\operatorname{wt}(\psi)$
determines which case occurs.
\end{enumerate}

\begin{lemma}\label{lem:app-recognition}
For $0\le i\le n$ and every stack $\psi$,
$E_i\psi\xrightarrow{w}\psi$ if and only if $w\in L_i$; in particular,
the parser accepts exactly $L_n$. Moreover, if
$\sigma\xrightarrow{w}\theta$ and $w\in L_i$, then $\sigma=E_i\theta$.
\end{lemma}
\begin{proof}
A run $E_i\psi\xrightarrow{w}\psi$ falls under case~(a) of (P3) with
$\sigma_1=E_i$, since case~(b) would end at $\theta_1\psi\ne\psi$. So
$w=w_1w_2$ with $E_i\xrightarrow{w_1}\lambda$ and
$\psi\xrightarrow{w_2}\psi$, and $w_2=\lambda$ by (P2). Together with
the lifting in (P3), it suffices to show that $E_i\xrightarrow{w}\lambda$
iff $w\in L_i$. We induct on $i$. From $E_0$ only the $\mathtt a$-move
is possible, and $\lambda$ admits no further move. For $i\ge1$, the
first move reads $\lceil$ or $\lfloor$. After $\lceil$, the stack
$E_{i-1}\rceil$ must be emptied; as its target $\lambda$ excludes
case~(b), (P3) with $\sigma_1=E_{i-1}$ and $\psi=\rceil$ splits the
remaining input as $zz'$ with $E_{i-1}\xrightarrow{z}\lambda$ and
$\rceil\xrightarrow{z'}\lambda$, whence $z'=\rceil$. After $\lfloor$,
two such splittings of $E_{i-1}E_{i-1}\rfloor$ give
$z_1z_2\rfloor$ with $E_{i-1}\xrightarrow{z_j}\lambda$. By induction,
$E_i\xrightarrow{w}\lambda$ iff
$w\in\lceil{}L_{i-1}\rceil\cup\lfloor{}L_{i-1}L_{i-1}\rfloor=L_i$, the
required runs being obtained by lifting.

For the last claim, the first part gives
$E_i\theta\xrightarrow{w}\theta$, and (P1) gives $\sigma=E_i\theta$.
\end{proof}

\begin{lemma}\label{lem:app-complete-subtree-exchange}
If $0\le i\le n$ and $w,w'\in L_i$, then $w\equiv_{L_n}w'$.
\end{lemma}
\begin{proof}
If $pwq\in L_n$, Lemma~\ref{lem:app-recognition} gives a run
$E_n\xrightarrow{p}\sigma\xrightarrow{w}\theta\xrightarrow{q}\lambda$
with $\sigma=E_i\theta$. The same lemma gives
$E_i\theta\xrightarrow{w'}\theta$, so $pw'q\in L_n$. Exchanging
$w$ and $w'$ proves the converse.
\end{proof}

\begin{lemma}[branch synchronization]\label{lem:app-branch-synchronization}
Let $\sigma\xrightarrow{x}\theta$ and $\sigma\xrightarrow{y}\theta$,
where $x,y$ are nonempty and begin with different letters. Then
$\sigma=E_i\psi$ for some $1\le i\le n$ and some stack $\psi$, and
$x=wx'$, $y=w'y'$ with $w,w'\in L_i$,
$\psi\xrightarrow{x'}\theta$ and $\psi\xrightarrow{y'}\theta$.
\end{lemma}
\begin{proof}
Only a top $E_i$ with $i\ge1$ admits two different letters, namely
$\lceil$ and $\lfloor$. Say $x=\lceil x_0$ and $y=\lfloor y_0$, so
$E_{i-1}\rceil\psi\xrightarrow{x_0}\theta$ and
$E_{i-1}E_{i-1}\rfloor\psi\xrightarrow{y_0}\theta$. Apply (P3) to both
runs with this $\psi$. Since the comparison of
$\operatorname{wt}(\theta)$ with $\operatorname{wt}(\psi)$ decides the
case, both runs are in the same case. Case~(b) is impossible:
$\theta=\theta_1\psi=\theta_2\psi$ would give $\theta_1=\theta_2$,
although $\theta_1$ ends with $\rceil$ and $\theta_2$ with $\rfloor$.
In case~(a), $x_0=x_1x'$ with $E_{i-1}\rceil\xrightarrow{x_1}\lambda$
and $\psi\xrightarrow{x'}\theta$. Thus $w:=\lceil x_1$ satisfies
$E_i\xrightarrow{w}\lambda$, and $w\in L_i$ by
Lemma~\ref{lem:app-recognition}. The same argument gives $y=w'y'$.
\end{proof}

\begin{proposition}\label{prop:app-path-congruence}
If $\sigma\xrightarrow{x}\theta$ and $\sigma\xrightarrow{y}\theta$, then
$x\equiv_{L_n}y$.
\end{proposition}
\begin{proof}
Induct on $|x|+|y|$. If $x$ or $y$ is empty, then $\sigma=\theta$, and
the other is empty by (P2). If $x=cx'$ and $y=cy'$ for a letter $c$,
determinism gives $\sigma\xrightarrow{c}\sigma'$ with
$\sigma'\xrightarrow{x'}\theta$ and $\sigma'\xrightarrow{y'}\theta$, so
$x'\equiv_{L_n}y'$ by induction and hence $x\equiv_{L_n}y$. Otherwise
Lemma~\ref{lem:app-branch-synchronization} gives $x=wx'$ and $y=w'y'$
as there. Since $w,w'$ are nonempty, induction gives
$x'\equiv_{L_n}y'$, Lemma~\ref{lem:app-complete-subtree-exchange} gives
$w\equiv_{L_n}w'$, and $\equiv_{L_n}$ is a monoid congruence.
\end{proof}

\begin{proof}[Proof of Proposition~\ref{prop:typed-thickness-gap}(i)]
The finite language $L_n$ is context-free. Let
$x,y\in\Gamma_{\mathrm{br}}^+$ and
$(u,v)\in\D_{L_n}(x)\cap\D_{L_n}(y)$. By
Lemma~\ref{lem:app-recognition} there are runs
$E_n\xrightarrow{u}\sigma\xrightarrow{x}\theta\xrightarrow{v}\lambda$
and
$E_n\xrightarrow{u}\sigma'\xrightarrow{y}\theta'\xrightarrow{v}\lambda$.
By (P1), $\sigma=\sigma'$ and $\theta=\theta'$, so
Proposition~\ref{prop:app-path-congruence} gives $x\equiv_{L_n}y$, that
is, $\D_{L_n}(x)=\D_{L_n}(y)$. Thus $L_n$ is Clark--Eyraud
substitutable. Adding the premise $s(x)=s(y)$ only weakens this
condition, so $L_n$ is $\sim_s$-substitutable for every $s$; in
particular $L_n\in\Ccf{s_0}\cap\Ccf{s_{\mathrm U}}$.
\end{proof}

\section*{Declaration of generative AI and AI-assisted technologies in the manuscript preparation process}
During the preparation and revision of this manuscript, the author used
OpenAI's ChatGPT and Anthropic's Claude to assist with language editing,
manuscript organization, and the examination of mathematical arguments.
The author reviewed and edited the AI-assisted content and takes full
responsibility for the results and proofs presented in this article.

\end{document}